\documentstyle[epsfig]{mn}

\newcommand\lta{\mathrel{\rlap{\lower 3pt\hbox{$\mathchar"218$}}
     \raise 2.0pt\hbox{$\mathchar"13C$}}}
\newcommand\gta{\mathrel{\rlap{\lower 3pt\hbox{$\mathchar"218$}}
     \raise 2.0pt\hbox{$\mathchar"13E$}}}
\newcommand\kms{km~s$^{-1}$}
\newcommand\kmsM{km~s$^{-1}\,$Mpc$^{-1}$}
\newcommand\etal{{et~al.}} 
\newcommand\sigth{\ifmmode \sigma_{\rm th}\else$\sigma_{\rm th}$\fi}
\newcommand\sigv{\ifmmode \sigma_v\else$\sigma_v$\fi}
\newcommand\mM{\ifmmode(m{-}M)\else$(m{-}M)$\fi}
\newcommand\msun{\ifmmode{\hbox{M$_\odot$}}\else{M$_\odot$}\fi}
\newcommand\iras{{\it IRAS}}
\newcommand\bi{\ifmmode \beta_{I}\else$\beta_I$\fi}
\newcommand\bo{\ifmmode \beta_{O}\else$\beta_O$\fi}
\newcommand\mgii{\ifmmode\hbox{Mg}_2\else{Mg$_2$}\fi}
\newcommand\hst{{\it HST}}
\def\PD{\hbox{\sc pd}}
\def\hkpc{$h^{-1}\,$kpc}
\def\cz{$c{z}$}
\def\mi{\ifmmode\overline{m}_I\else$\overline{m}_I$\fi}
\def\mv{\ifmmode\overline{m}_V\else$\overline{m}_V$\fi}
\def\mbar{\ifmmode\overline{m}\else$\overline{m}$\fi}
\def\Mbar{\ifmmode\overline{M}\else$\overline{M}$\fi}
\def\lbar{\ifmmode\overline{L}\else$\overline{L}$\fi}
\def\ibar{\ifmmode\overline{I}\else$\overline{I}$\fi}
\def\vbar{\ifmmode\overline{V}\else$\overline{V}$\fi}
\def\vbib{\ifmmode(\overline{V}{-}\overline{I})\else$(\overline{V}{-}\overline{I})$\fi}
\def\vbkb{\ifmmode(\overline{V}{-}\overline{K})\else$(\overline{V}{-}\overline{K})$\fi}
\def\ibkb{\ifmmode(\overline{I}{-}\overline{K})\else$(\overline{I}{-}\overline{K})$\fi}
\def\Mi{\ifmmode\overline{M}_I\else$\overline{M}_I$\fi}
\def\Miz{\ifmmode\overline{M}_I^0\else$\overline{M}_I^0$\fi}
\def\Mv{\ifmmode\overline{M}_V\else$\overline{M}_V$\fi}
\def\vi{\ifmmode(V{-}I)\else$(V{-}I)$\fi}
\def\viz{\ifmmode(V{-}I)_0\else$(V{-}I)_0$\fi}

\def\dn{\ifmmode D_n\hbox{-}\sigma\else$D_n\hbox{-}\sigma$\fi}
\def\Nbar{\ifmmode\overline{N}\else$\overline{N}$\fi}
\def\ho{\ifmmode H_0\else$H_0$\fi}
\def\lsig{\ifmmode \log(\sigma)\else$\log(\sigma)$\fi}
\def\xfp{\ifmmode X_{\rm FP}\else$X_{\rm FP}$\fi}
\def\sbe{\ifmmode \langle\mu\rangle_e \else $\langle\mu\rangle_e$\fi}
\def\aj{AJ}
\def\apj{ApJ}
\def\apjl{ApJ}
\def\apjs{ApJS}
\def\mnras{MNRAS}

\def\pasp{PASP}
\def\aap{A\&A}

\title[FP and SBF Distances]{Early-type Galaxy Distances from the Fundamental Plane and Surface Brightness Fluctuations}

\author[J.\ P.\ Blakeslee et al.]{John P. Blakeslee,$^{1,2}$
John R.\ Lucey,$^{2}$ John L.\ Tonry,$^{3}$ Michael J.\ Hudson,$^{4}$ 
\newauthor Vijay K.\ Narayanan,$^{5}$ and Brian J.\ Barris$^{3}$ \\
$^1${Department of Physics and Astronomy,
Johns Hopkins University, Baltimore, MD 21218, U.S.A.; jpb@pha.jhu.edu}\\
$^2${Department of Physics, University of Durham, South Road, 
Durham, DH1\,3LE, United Kingdom; John.Lucey@durham.ac.uk}\\
$^3${Institute for Astronomy, University of Hawaii,
2680 Woodlawn Drive, Honolulu, HI 96822, U.S.A.; jt,barris@ifa.hawaii.edu}\\
$^4${Department of Physics, 
University of Waterloo, ON, N2L\,3G1, Canada; mjhudson@uwaterloo.ca}\\
$^5${Department of Astrophysical Sciences, Princeton University,Princeton, NJ 08544; vijay@astro.princeton.edu}
}

\date{Accepted --- 2001. Received --- ; in original form --- }
\pagerange{\pageref{firstpage}--\pageref{lastpage}}\pubyear{2001}

\begin{document}

\maketitle \label{firstpage}

\begin{abstract}
We compare two of the most popular methods for deriving distances to
early-type galaxies: the fundamental plane (FP) and surface brightness
fluctuations (SBF).  Distances for 170 galaxies are compared.  
A third set of distances is provided by predictions
derived from the density field of the \iras\ redshift survey.
Overall there is good agreement between the different distance indicators.
We investigate systematic trends in the residuals of the three
sets of distance comparisons. 
First, we find that several nearby, low-luminosity, mainly S0 galaxies
have systematically low FP distances.  Because these galaxies
also have \mgii\ indices among the lowest in the sample, we conclude
that they deviate from the FP partly because of recent star
formation and consequently low mass-to-light ratios; differences
in their internal velocity structures may also play a role.
Second, we find some
evidence that the ground-based $I$-band SBF survey distances (Tonry \etal\
2001) begin to show a bias near the survey limit at $c{z} \gta 3500$ \kms,
which is expected for this sort of distance-limited survey, but had not
previously been demonstrated.  
Although SBF and FP distances are affected in opposite senses by errors
in the Galactic extinction estimates, we find no
evidence for biases in the distances due to Galactic extinction.
The tie between the Cepheid-calibrated SBF distances (Mpc) and the far-field
calibrated FP distances (\kms) yields a Hubble constant of $H_0 =
68\pm3$ \kmsM, while the comparison between SBF and the IRAS-reconstructed
distances yields $H_0 = 74\pm2$ \kmsM\ (independent errors only).  Thus,
there is a marginal inconsistency in the
direct and \iras-reconstructed ties to the Hubble flow (this can be seen
independently of the SBF distances).  
Possible explanations include systematic errors in the
redshift survey completeness estimates or in the FP aperture corrections,
but at this point, the best
estimate of \ho\ may come from a simple average of the above two estimates.
After revising the SBF distances downward by 2.8\% to be in agreement with the
final set of Key Project Cepheid distances (Freedman \etal\ 2001), we
conclude $H_0=73\pm4\pm11$ \kmsM\ from early-type galaxies,
where the second errorbar represents the total systematic uncertainty in 
the distance zero~point.
We also discuss the `fluctuation star count' $\Nbar\equiv \mbar-m_{\rm tot}$,
recently introduced by Tonry \etal\ (2001) as a less demanding alternative
to \vi\ for calibrating SBF distances.  The $\Nbar$-calibrated SBF method
is akin to a hybrid SBF-FP distance indicator, and we find that the
use of \Nbar\ actually improves the SBF distances.  Further study of the 
behavior of this quantity may provide an important new test for
models of elliptical galaxy formation.
\end{abstract}
\begin{keywords}
galaxies: distances and redshifts --- galaxies: elliptical and lenticular, cD  ---
galaxies: fundamental parameters --- galaxies: stellar content
\\ 
\end{keywords}

\section{Introduction}

Early attempts to gauge relative distances in samples of elliptical
galaxies were based primarily on the colour-magnitude (de Vaucouleurs 1961;
Sandage 1972; Sandage \& Visvanathan 1978) and Faber-Jackson relations
(Faber \& Jackson 1976; Schechter 1980; Tonry \& Davis 1981), which
treated ellipticals as a structurally and chemically homogeneous,
single-parameter family.  The introduction of the two-parameter
fundamental plane (FP), and the closely related \dn\ relation, by Dressler
\etal\ (1987) and Djorgovski \& Davis (1987) allowed for a significant
improvement in the measurement of early-type distances.  Since then, there
has been a proliferation of FP studies, concentrating on the early-type
populations within rich clusters (e.g., Lucey \& Carter 1988; Lucey \etal\
1991a,b,c; Jorgensen \etal\ 1993, 1996; Prugniel \& Simien 1996; Hudson
\etal\ 1997; Pahre \etal\ 1998; Colless \etal\ 2001).  Although FP
measurements have been made at redshifts $z{\,\approx\,}0.5$ for galaxy
evolution studies (van Dokkum \& Franx 1996; Pahre 1998; Kelson \etal\
2000b), individual cluster peculiar velocities are only measurable at
$z{\,\lta\,}0.05$.

The streaming motions of Abell clusters (SMAC) project (Smith \etal\ 2000,
2001, hereafter SMAC-I and SMAC-II; Hudson \etal\ 2001, hereafter
SMAC-III) is an FP survey with a limiting depth of 12,000 \kms. It combines
literature data from $\sim\,$20 different sources with an extensive set of
new photometric and spectroscopic observations.  A great deal of effort
was made in standardizing the data sets to a homogeneous system through
intercomparison of data for overlapping galaxies.  As a result, the SMAC
data set is the largest high-precision FP data set currently available.

In comparison to the FP, the surface brightness fluctuations (SBF) method
developed by Tonry \& Schneider (1988) has employed a smaller number of
researchers.  The vast majority of the extant SBF data were collected as
part of the ground-based $I$-band SBF survey described in detail by Tonry
\etal\ (1997, hereafter SBF-I).  This includes the early studies by Tonry,
Ajhar, \& Luppino (1990) and Tonry (1991), although the results from these
early papers have been significantly revised on account of improved
analysis methods, better data and standardization of the photometry, new
extinction estimates, and a new calibration of the dependence of SBF on
the stellar population.  The full SBF survey data set, comprising about
300 galaxies within $c{z}\lta4000$ \kms, is presented by Tonry \etal\
(2001, hereafter SBF-IV).  In addition, a number of recent studies have
taken advantage of the superior resolution of the {\it Hubble Space
Telescope} (\hst) for SBF measurements (Ajhar \etal\ 1997; Lauer \etal\
1998; Neilsen \& Tsvetanov 2000; Jensen \etal\ 2001) reaching as far as
10,000~\kms.  A recent comprehensive review of SBF is
given by Blakeslee, Ajhar, \& Tonry (1999a).

In this paper, we perform a series of comparisons between the SBF and FP
distance methods.  The following section describes the SBF and FP
calibrations, distances, and Malmquist corrections, as well as the hybrid
`\Nbar\,SBF' distance estimator and distances predicted using the
line-of-sight distance-redshift relations derived from the observed \iras\
galaxy density field.
The FP distances use photometric parameters derived by
Blakeslee \etal\ (2001a, hereafter Paper~I) from the SBF 
survey data images and velocity dispersions from the SMAC survey.
Section~3 performs a series of direct
comparisons between the distances obtained by the different methods for
individual galaxies.  Section~4 discusses potential systematic biases in
the distance methods, particularly in regard to the Hubble flow, or
alternatively, the value of the Hubble constant, implied by the early-type
galaxy distance scale.  The final section summarizes our main results.

\section{The Distances}
\label{sec:distances}
 The present work intercompares three types of galaxy distances: SBF
 measurements, FP measurements, and distances predicted from the observed
 galaxy density field.  The predicted distances are used
 mainly for investigating potential problems in the SBF-FP comparison.  
We give an overview of these three types of distance estimation
 before directly comparing the distances.

\subsection{SBF Distances}
\subsubsection{\vi-calibrated SBF}
\label{ssec:sbfdistances}

We use the data from the ground-based SBF survey as
tabulated by SBF-IV.  The distances are calculated from
the measured $I$-band apparent fluctuation magnitude \mi\
and the \viz\ colour according to
\begin{eqnarray}
\overline M_I \;=\; -1.74 \,+\, 4.5\,[(V{-}I)_0 - 1.15\,] \label{eq:vicalib} \\
(m-M) \;=\; (\mi - \Mi)\; \label{eq:mMidentity}
\end{eqnarray}
(Tonry \etal\ 2000, hereafter SBF-II).
The colour dependence is based on the behavior of \mi\ among
galaxies in groups, and the zero point is based on SBF measurements in
the bulges of six spirals for which Cepheid distances have been
measured.  Problems and uncertainties in this empirical calibration are
discussed in Appendix~B of SBF-II.  The Cepheid distances are
those tabulated by Ferrarese \etal\ (2000) for the \hst\ Key Project
on the Distance Scale. 
The total uncertainty in the \Mi\ zero point is about 0.2~mag and is
dominated by systematic effects, such as the LMC distance.  For a
discussion of the theoretical SBF calibration and uncertainty from stellar
population modeling, see Liu \etal\ (2000) and Blakeslee \etal\ (2001b).
 
Freedman \etal\ (2001) have recently updated
the Key Project distances using the revised
Cepheid P-L relation from Udalski \etal\ (1999) (although keeping
the LMC distance modulus at 18.50\,mag) and the
metallicity correction from Kennicutt \etal\ (1998).
The effects of these two changes on the distances to the
SBF calibrating spirals nearly cancel.  The resulting SBF 
direct calibration would be about 0.06 mag fainter, well
within the uncertainty of~Eq.\,(\ref{eq:vicalib}).  

We adopt the SBF distance errors from SBF-IV, which are 0.22~mag in the
median, but we add an allowance for uncertainty in the Galactic extinction
values from Schlegel, Finkbeiner, \& Davis (1998, hereafter SFD).
This is important because extinction errors cause the SBF and FP
distances to go in different directions.
Because of the steep colour-dependence of \Mi, an underestimate of the true
extinction produces an underestimate of the SBF distance:
$\pm\delta\mM_{\rm SBF} = {\pm}3.8\,\delta E(B{-}V)$, where we have used
Eq.\,(\ref{eq:vicalib}) and the extinction ratios recommended by SFD. For
the FP and most other distance indicators, underestimating the extinctions
produces overestimated distances: $\pm\delta\mM_{\rm FP} = {\mp}2.6\,\delta E(B{-}V)$,
in the $R$~band.
SFD concluded that the error on a given extinction value was 16\% of the value,
i.e., $|\delta E(B{-}V)| =  0.16\,E(B{-}V)$. 
This implies a distance error from extinction of $0.61\,E(B{-}V)$ for
\viz-calibrated SBF and $0.42\,E(B{-}V)$ in the opposite sense for
$R$-band FP, or $\sim\,$1.0$\,E(B{-}V)$ for the difference.
Therefore, the agreement will be worse in areas of high extinction;
conversely, it is usually possible to bring the FP and SBF distances
for any given galaxy at high extinction into agreement by changing the
extinction estimate.

\subsubsection{\Nbar-calibrated SBF}
SBF-IV introduced an alternative calibration for SBF based on 
the distance-independent fluctuation count \Nbar, defined as
\begin{equation}
\Nbar \,=\, \mbar - m_{\rm tot} \,=\,
    +2.5\,\log\left[L_{\rm tot} \over \lbar\right] \,,
\label{eq:nbardef}
\end{equation}
where $m_{\rm tot}$ and $L_{\rm tot}$ are the total apparent 
magnitude and luminosity of a galaxy and \lbar\ is the luminosity
corresponding to magnitude \mbar\ at the distance of the galaxy.
SBF-IV derived the following empirical
calibration based on this new parameter:
\begin{equation}
  \Mi \,=\, -1.74 + 0.14(\Nbar-20.0) \,.
  \label{eq:nbarmbar}\\
\end{equation}
The distance modulus is then given by
\begin{equation}
 (m-M) \;=\; 0.86\,\mi \,+\, 0.14\,m_{{\rm tot},I} \,+\, 4.54 \,. \label{eq:mMnbar}
\end{equation}
Thus, the distance moduli are 14\% less sensitive to errors
in \mi\ and the goal of a $\lta\,$0.02 mag uncertainty in \vi\
is replaced by that of a $\lta\,$0.6 mag uncertainty in the total magnitude.
Moreover, the moduli are about 50\% less sensitive to the
Galactic extinction and are affected in the sense opposite
to how they are affected when using the \viz\ calibration
of Eq.\,(\ref{eq:vicalib}).

However, use of \Nbar\ removes one of the most attractive features
of SBF, its pure basis in stellar populations.  The validity of the
calibration relies on the universality of early-type galaxy scaling
relations, in particular the mass-metallicity relation 
(e.g., Guzm\'an \etal\ 1992; Bender, Burstein, \& Faber 1993).
Although Blakeslee \etal\ (2001b) showed how composite population 
models combined with scaling relations can produce a tight relation
involving \mbar\ and $m_{\rm tot}$, the \Nbar-based SBF method
cannot be calibrated in a purely theoretical way.

\begin{figure}
\medskip\vbox{\centering\leavevmode\hbox{
\epsfxsize=8.0cm
\epsffile{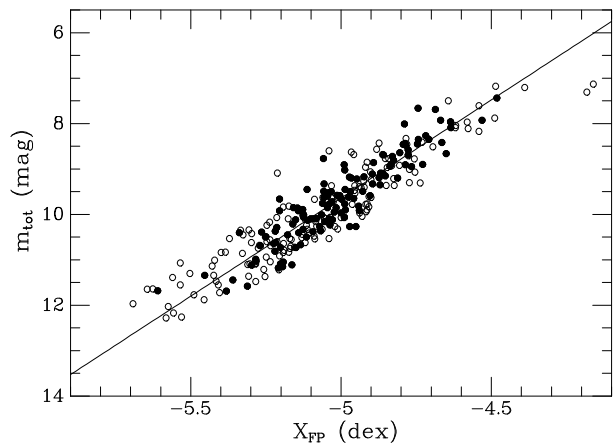}
}\caption{\small
Illustration of the covariance between the estimated total apparent
magnitude $m_{\rm tot}$ and the FP photometric parameter 
$\xfp\equiv\log R_e - 0.33\sbe$. Filled
circles are for $T{\,=\,}{-}5$ ellipticals; open circles are other
galaxies. The solid line is a least-squares fit to the ellipticals;
the slope is 4.3.
The two open circles falling below the line at the extreme right
are the Local Group compact elliptical M32 and the
nearby spiral M81, neither of which are used for any
of the distance comparisons below.
\label{fig:mtot_xfp}}}
\end{figure}

Most importantly for the present work, the use of \Nbar\ introduces strong
covariance between the SBF and FP distances, as the value of $m_{\rm tot}$
used for \Nbar\ is strongly correlated with the FP photometric combination
$\xfp\equiv \log R_e - \beta\sbe$ (see \S\ref{ssec:fpdist}).
Figure~\ref{fig:mtot_xfp} shows the correlation: at fixed \mi, the SBF
distance modulus determined from \Nbar\ goes as $\sim 0.6\,\xfp $.  This
covariance produces a tight correlation and a non-unit slope for the
relation between \Nbar-derived $\mM_{\rm SBF}$ and $\mM_{\rm FP}$, making
the results difficult to interpret.  SBF-IV and Paper~I also showed good
correlations between \Nbar\ and $\log\sigma$, which essentially reflects
the scaling of velocity dispersion with the total number of stars in a
galaxy.  Although comparisons between `\Nbar\,SBF' and FP distances are
therefore not meaningful, we {\it can} compare the \Nbar\,SBF distances to
those predicted from the \iras\ density field and see how the agreement changes
with respect to the standard \viz-calibrated SBF distances.

\subsection{FP distances}
\label{ssec:fpdist}

The FP relation can be represented as
\begin{equation}
\log R_e \;=\; \alpha \log \sigma \,+\, \beta \sbe \,+\, \gamma \,.
 \label{eq:fprel}
\end{equation}
If the distance is known by other means, so that $R_e$ is measured in
physical units (kpc) then $\gamma$ is a constant.  More commonly,
$R_e$ is measured in angular units, so that determination of $\gamma_i$
for an individual galaxy or cluster yields the angular diameter distance
$d_{A,i}$ relative to some calibrating cluster of known angular diameter
distance $d_{A,0}$ and zero~point $\gamma_0$:
\begin{equation}
d_{A,i} \;=\; d_{A,0}\, 10^{(\gamma_0 - \gamma_i)} \,.
 \label{eq:fpdist}
\end{equation}
One then solves the angular distance equation (we use $q_0{\,=\,}0$,
but this is unimportant for the present sample)
to obtain the Hubble distance, i.e., the redshift $c{z}$ the galaxy
would have if its peculiar velocity were zero.

When $\gamma_i$ is determined from a fit that minimizes the residuals in
the distance-dependent quantity $\log R_e$, the method is known as the
`forward' FP; when residuals are minimized in $\log \sigma$, it is the
`inverse' FP.  Of course, it is also possible to perform a fit
minimizing the scatter orthogonal to the plane (e.g., Jorgensen \etal\
1996).  For any given galaxy, the forward FP provides the best estimate of
the distance, as long as the fit coefficients have been determined from
an unbiased sample (which is the difficult part).  The inverse FP reduces
selection biases for magnitude- or diameter-limited cluster galaxy
samples at different distances (Schechter 1980), 
but does not yield the `best distance' for any given galaxy.
Strauss \& Willick (1995) give a detailed discussion of the relative
merits of the forward and inverse methods.

Colless \etal\ (2001) tabulate values of $\alpha$ and $\beta$
(actually $B\equiv2.5\beta$) from 9 different FP studies over
the last decade.  The values of $\beta$ show good consistency,
having a range from 0.316 to 0.348, with an average of 0.326
and a dispersion less than 0.01.  The SMAC survey determined
a best fit value $\beta=0.338$.  We fix
$\beta=0.33$, or $B=0.825$, as used for the \xfp\ comparisons
of Paper~I.  This is very close to the coefficients
derived from very large data samples by Hudson \etal\ (1997),
Jorgensen \etal\ (1996), Colless \etal\ (2001) and the SMAC project.

The best-fitting $\alpha$ will depend on whether one does a forward,
inverse, or orthogonal fit to the FP.  The SMAC survey used the inverse
FP and derived $\alpha=1.42$.  Forward and orthogonal fits give smaller
values; for example, Jorgensen \etal\ (1996) obtained
$\alpha=1.24\pm0.07$.  Although we are using SMAC data, we are interested
in comparing FP and SBF distances to individual galaxies, unlike SMAC,
which was a cluster survey.  Thus, we use the forward form of the
relation derived from the Coma cluster, for which the SMAC data set is
about twice as large as for any other cluster. 

\begin{figure}
\medskip\vbox{\centering\leavevmode\hbox{
\epsfxsize=8.0cm\epsffile{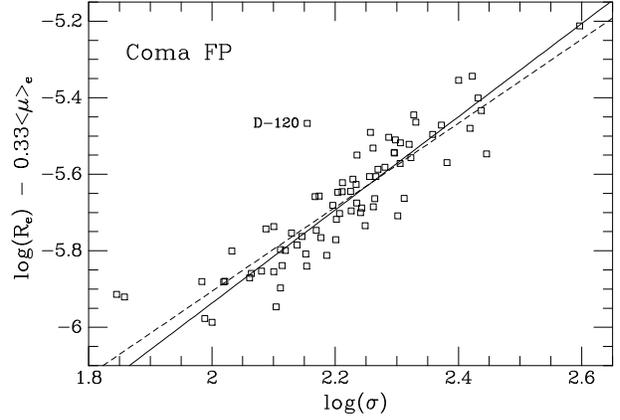}
}\caption{\small
The `forward' $R$-band FP for galaxies in the Coma cluster with SMAC data.
The effective radii $R_e$ are in arcseconds, the surface brightness
\sbe\ are in mag~arcsec$^{-2}$, and the velocity dispersions $\sigma$
are in \kms.
The dashed line is a least-squares fit to the full sample of 76 galaxies;
its slope of 1.10 is too shallow because of incompleteness at
small values of \xfp\ (\xfp\ is biased high 
at small values of the velocity dispersion $\sigma$ because of
selection effects).  
The solid line is a fit to the 54 galaxies with $\log\sigma\ge2.15$ and
has a slope of 1.22.  This fit is used to define the forward SMAC FP. 
The galaxy D-120 (labelled) was omitted from the fit, but the slope
is not very sensitive to this.
\label{fig:ComaFP}}}
\end{figure}

Because of the selection effect called `diameter bias' by Lynden-Bell
\etal\ (1988), the slope $\alpha$ of the forward fit to a
magnitude-limited sample will be too shallow and the zero point
too high.  Cutting the sample at progressively higher values of
$\log\sigma$ yields steeper slopes up to $\log\sigma\sim2.1$, at which
point $\alpha=1.22\pm0.09$.  Figure~\ref{fig:ComaFP} shows the relation
and a fit to the data with $\log\sigma\ge2.15$ ($\sigma_{\rm
cut}{\,=\,}140$ \kms) and the moderate outlier D-120 excluded.  The rms
scatter is 0.064~dex (0.073 dex including D-120) in \xfp, or 0.32--0.37
mag, implying a $\sim 17$\% distance error per galaxy.  The FP distance
error for cluster galaxies is typically about 20\% (see the compilation
by Colless \etal\ 2001).  For the adopted $\alpha=1.22$ and
$\log\sigma>2.15$, the $R$-band Coma FP zero point is $\gamma_{\rm
Com}{\,=\,}{-}8.377\pm0.009$.

Finally, we note that these values of $\alpha$ are measured in the
$R$~band, and $\alpha$ may depend on the bandpass.
The main evidence for this comes from the result
$\alpha=1.53\pm0.08$ found by Pahre \etal\ (1998) from an orthogonal fit
to the $K$-band FP, as compared to the significantly shallower slope
found by Jorgensen \etal\ (1996) and others. However,
the data samples and analyses methods also differ among these studies.
In contrast, Girardi \etal\ (in preparation)
have derived the FP of a homogeneous sample of 9000 galaxies in the
Sloan $g^\prime r^\prime i^\prime z^\prime$ bands and find no
significant variation in $\alpha$ among these bands for a given type of
FP fit.

\subsection{Inhomogeneous Malmquist Corrections}
\label{ssec:malmquist}

Malmquist (1920) investigated the luminosity bias in flux-limited samples
and the resulting distance error which occurs near the sample limit when
assuming all the objects have the same luminosity.  Here, we use
`Malmquist bias' as in the terminology of Lynden-Bell \etal\ (1988) to
refer to the distance bias arising from the spatial distribution of the
sample galaxies, including the increase in the volume element with
distance.  We note that this differs from the selection biases referred
to by the same name by Schechter (1980), Aaronson \etal\ (1986), and
others (see Strauss \& Willick 1995 for details).
By `inhomogeneous Malmquist bias' we mean the apparent tendency of
galaxies to be scattered out of regions of higher density by the
observational errors.  As shown by Hudson (1994), correcting for this bias
is especially important when treating early-type galaxies as field
galaxies.  In particular, Hudson found that inhomogeneous Malmquist bias
was largely responsible for previous claims of `backside' infall into
the Great Attractor (e.g., Dressler \& Faber 1990).

We essentially follow Strauss \& Willick (1995) in correcting for
inhomogeneous Malmquist bias. This approach forms the basis of the VELMOD
maximum likelihood method used by Willick \etal\ (1997) and Willick \&
Strauss (1998) to constrain the properties of local mass density field
from galaxy peculiar velocities measured with the Tully-Fisher
method.  In particular, they constrained the density parameter
$\beta_I\equiv\Omega^{0.6}/b_I$, where $b_I$ is the linear biasing factor
of the \iras\ 1.2\,Jy survey (Fisher \etal\ 1995) galaxies with respect to
the mass density field and $\Omega$ is the mean cosmic density.  The
VELMOD method was also applied to the SBF survey peculiar velocities by
Willick \& Batra (2001).

In the notation of Strauss \& Willick, an unbiased estimate of the
true distance $r$ for a galaxy of measured distance $d$ is given
by the expectation value
\begin{equation}
E(r|d) \;=\; { {\int_{0}^{\infty} r^3\, n(r)\,
     \exp\left\{ -{[\ln(r/d)]^2 \over 2\Delta^2} \right\}\, d{r} }
\over  {\int_{0}^{\infty} r^2\, n(r)\,
     \exp\left\{ -{[\ln(r/d)]^2 \over 2\Delta^2} \right\} \, d{r} }} \,,
\label{eq:epnval}
\end{equation}
where n(r) is the real-space galaxy density distribution in the
direction of the given sample galaxy, $\Delta$ is the fractional
distance error, and the errors are assumed Gaussian in $\log$ distance.
Note that because it goes as $n(r)r^2$, the distance probability tends
to have a tail to larger distances.  This means that the unbiased
distance $E(r|d)$ tends to be larger than the maximum-probability 
distance, which is usually closer to the measured~$d$.
For the case of a uniform density, Eq.\,(\ref{eq:epnval}) yields the
Lynden-Bell \etal\ (1988) homogeneous Malmquist correction factor of
$\,e^{\,3.5 \Delta^2}$.

We estimate $n(r)$ and derive the redshift-distance relation along the
line of sight to each of our sample galaxies using the redshift
distribution of the \iras~1.2 Jy survey under the assumptions of linear
theory, a linear biasing model with $\beta_I=0.4$, and a power-preserving
filter with a smoothing scale of 300 \kms\ (see Willick \etal\ 1997 and
Willick \& Strauss 1998 for details). 
The comparisons of the Malmquist-corrected FP and SBF distances are not
very sensitive to the value of $\beta_I$ adopted.  However, Blakeslee
\etal\ (1999b, hereafter SBF-III), using the spherical harmonic expansion
method of Nusser \& Davis (1994), found that the comparison between SBF
survey peculiar velocity measurements and the \iras\ density field
predictions implied $\beta_I=0.42^{+0.10}_{-0.06}$, and Riess \etal\
(1997) found $\beta_I = 0.40\pm0.15$ using the same method but SNIa
peculiar velocities.  Willick \& Batra (2001), whose methods and `IRAS
model' we have largely adopted for this analysis, found
$\beta_I=0.38\pm0.06$, again using SBF survey peculiar velocities.

One difference with respect to Willick \& Batra
is that we do not directly insert $n(r)$ determined from the \iras\
galaxies into Eq.\,(\ref{eq:epnval}), where the relevant density 
distribution is that of the early-type galaxies in our sample.
Instead, we write the early-type galaxy density as
\begin{equation}
n_e(r) \;=\; 1 \,+\, {b_e \over b_I}\, \delta_I(r) \,,
\label{eq:ne}
\end{equation}
where $\delta_I(r)$ is the relative overdensity of the \iras\ galaxies and
$b_e$ and $b_I$ are the biasing factors of the early-type and
\iras\ galaxies, respectively.
Baker \etal\ (1998) found the relative biasing between optically-selected
($R$ band) and \iras\ galaxies to be ${b_o \over b_I} \approx 1.4$, while
Willmer \etal\ (1998) and Narayanan, Berlind, \& Weinberg (2000) found 
$\frac{b_e}{b_S} \approx 1.2$ for the relative biasing of early-types
and spirals.  If early-type galaxies make up 10--50\% of optically
selected galaxies, the relative biasing factor of early-type to
\iras\ galaxies is then 1.5--1.6.  We adopt ${b_e \over b_I} = 1.5$,
although the results of the analysis would change very little
for ratios in the range 1.0--1.7, and use
Eq.\,(\ref{eq:ne}) for the density in Eq.\,(\ref{eq:epnval}).

\subsection{Distances from the Density Field}

The peculiar velocity predictions from the observed galaxy
density field provide a third set of distances for the galaxies
in our sample.  These predictions are usually compared to the
observed peculiar velocities in order to constrain $\beta_I$,
and thus $\Omega$.  Here, we are mainly interested in having an
independent set of distances for evaluating the SBF--FP comparisons,
and so we simply adopt the \iras\ model described above.
Again, following Willick \& Batra (2001), the expectation value
of the true distance $r$ for an observed redshift \cz\ along a given
line of sight is
\begin{equation}
E(r|c{z}) \;=\;  {\int_{0}^{\infty} r^3\, {n(r) \over \sigma_v(r)}
 \exp\left\{ -{(c{z} - [r + u(r)])^2 \over 2\,\sigma^2_v(r) }\right\} d{r}
     \over  {\int_{0}^{\infty} r^2\, {n(r) \over \sigma_v(r)} 
 \exp\left\{ -{(c{z} - [r + u(r)])^2 \over 2\,\sigma^2_v(r)}\right\} d{r} }}\,,
\label{eq:epnrv}
\end{equation}
where \sigv\ is the thermal noise in the velocity field and $u(r)$
is the radial component of the predicted peculiar velocity.
We allow the thermal noise for this early-type SBF galaxy
sample to vary with local \iras\ galaxy density as
\begin{equation}
\sigma_v(r) \;=\; [185 \,+\, 15\,\delta_I(r)]\;\hbox{\kms} \,
\label{eq:sigvar}
\end{equation}
(Willick, private communication).  To estimate the error $\delta r$ in
the predicted distance, we calculate $E(r^2|c{z})$ analogously
to Eq.\,(\ref{eq:epnrv}) and then 
\begin{equation}
\delta r \;=\; \sqrt{E(r^2|cz) - [E(r|c{z})]^2} \,.
\label{eq:sigrpred}
\end{equation}
The actual errors in the \iras-predicted distances may be
greater than this in general, as the adopted model will not
be a perfect description of the density field and
non-linear effects can be significant.

In calculating the predicted distances, we 
use the observed galaxy velocities (transformed to the
Local Group frame) except in a few cases where 
there are known very large peculiar velocities.  
For Virgo, we set all 16 sample galaxies within $5^\circ$ of M87
and local group velocity $v_{\rm LG}<2700$ \kms\ to
have $v_{\rm LG} = 1035$ \kms\ and $\sigv = 50$ \kms.
The Eridanus group galaxy NGC\,1400 has a relative velocity of
$-$1215 \kms\ with respect to NGC\,1407, a larger elliptical
at a projected separation of only 63\,\hkpc.  We simply
change the velocity of NGC\,1400 to be that of NGC\,1407.
In Centaurus, a variety of evidence, including colour-magnitude relations
and luminosity functions (Lucey, Currie, \& Dickens 1986), the optical
$D_n$--$\sigma$ relation (Lucey \& Carter 1988), the near-infrared FP
(Pahre \etal\ 1998), and high resolution SBF observations from \hst\
Cycle~6 (discussed in SBF-II), indicates that the `Cen-30' and
`Cen-45' ellipticals lie at the same physical distance.  Thus, we
reassign the 3 Cen-45 galaxies in our sample to have $v_{\rm LG} = 2935$
\kms\ (for a CMB velocity $v_{\rm CMB} = 3500$ \kms).

\begin{figure}
\medskip\vbox{\centering\leavevmode\hbox{
\epsfxsize=8.0cm
\epsffile{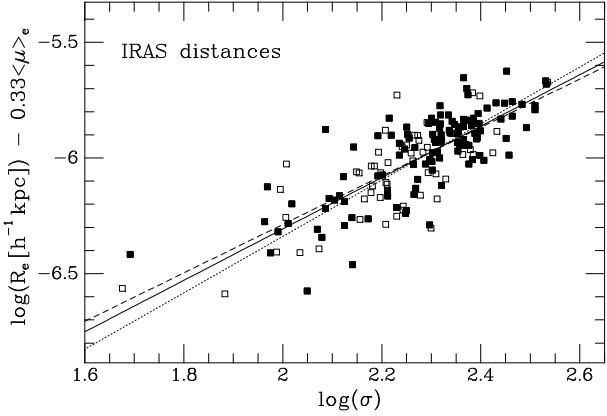}
}\caption{\small
The `forward' FP for galaxies in the present cross-matched
sample of SBF--SMAC galaxies.  The distances used in transforming
the $R_e$ values to a common scale (\hkpc) are those derived from the
\iras\ redshift survey density field as described in the text.
The dashed line (with a slope of 1.05) and the solid line (with slope 1.11)
show the linear fits to the full and
$\log\sigma{\,>\,}2.1$ samples, respectively, and the dotted
line shows a fit to the $\log\sigma{\,>\,}2.1$ sample for a
fixed slope of 1.22, as for the Coma FP.
Solid and open symbols represent ellipticals and S0s, respectively,
but the fits shown are to all galaxies regardless of type;
restricting the fit to ellipticals produces virtually identical results.
\label{fig:irasFPcal}}}
\end{figure}

These predicted distances and the SBF distances both could be used to derive
the FP coefficients from the sample galaxies directly, without
using the SMAC Coma cluster data.  For two reasons we have elected
not to do this.  First, we wish to look for possible systematic
problems or trends in the distance comparisons, and so it is best
to keep the different types of distances independent.  Second, we
are also interested in the value of \ho, so we wish to tie as 
directly as possible the Cepheid-calibrated distances (in Mpc)
from SBF to the Hubble flow-calibrated distances (\kms) from the
FP and \iras.  For illustration, however, Figure~\ref{fig:irasFPcal}
shows the \iras-calibrated FP of our sample galaxies.  
The best-fitting slope for the galaxies with $\log \sigma>2.1$
is $1.11\pm0.11$ ($1.05\pm0.07$ for the full sample), 
consistent with the Coma slope.  The rms scatter is 0.12 dex
(0.11 dex for the ellipticals),
which includes the errors in the estimated distances.  
However, the best agreement with the Coma zero point is achieved 
if Coma has a CMB-frame
peculiar velocity of about $+$1000 \kms.  We discuss the results of
the distance comparisons in greater detail below.

\section{Comparison of Distances}

We matched all galaxies beyond the Local Group having SBF distances 
(distance modulus uncertainty $<0.7$) against all galaxies with FP
distances constructed from SMAC velocity dispersions and SBF survey
photometry (Paper~I).  There are a total of 170 galaxies in this cross-matched
sample.  Of these, we omit 6 galaxies from the $\chi^2$ analyses for the
following reasons.  Three galaxies (NGC\,404, NGC\,4476, and
NGC\,4489) have velocity dispersions $\sigma<50$ \kms\ (no others
have $\sigma<70$ \kms) and thus would have very uncertain FP distances.
One galaxy (NGC\,3641, a close companion of the much brighter NGC\,3640)
has an FP distance over 7000 \kms,
even though its velocity is only $\sim2000$ \kms;
it is $\sim5.5\sigma$ outlier in both the FP-SBF and FP-\iras\ comparisons.
Two galaxies (NGC\,2305 and E322-059, also called NGC\,4645A) have 
exceptionally poor SBF data with quality parameters $\PD>3.6$ 
(see \S\ref{ssec:sbfselect} below for the definition of \PD) and
probably biased distance estimates as a result.
Both are $\sim3\sigma$ outliers in the FP-SBF comparisons, but 
NGC\,2305 is $5.8\sigma$ outlier in the SBF-\iras\ comparison because
it is in a region of the sky where the predicted error on the
\iras\ distance is very small (0.11~mag).
Note that this is the only data-quality cut we make (and we still plot
these galaxies in the comparison figures below), as we wish
to explore the possibility of bias in the poorest quality SBF survey data,
an issue raised in SBF-II where a much more severe \PD\ cut was made
to avoid any potential selection bias.  We are left with 164 galaxies
for judging goodness of fit for the distance comparisons.  Of the
6 rejected, only 2 are big outliers (NGC\,3641 in the FP-SBF and
FP-\iras\ comparisons and NGC\,2305 in the SBF-\iras\ comparison), although
all six are strongly suspected of systematic problems.

\begin{figure}
\medskip\vbox{\centering\leavevmode\hbox{
\epsfxsize=8.0cm
\epsffile{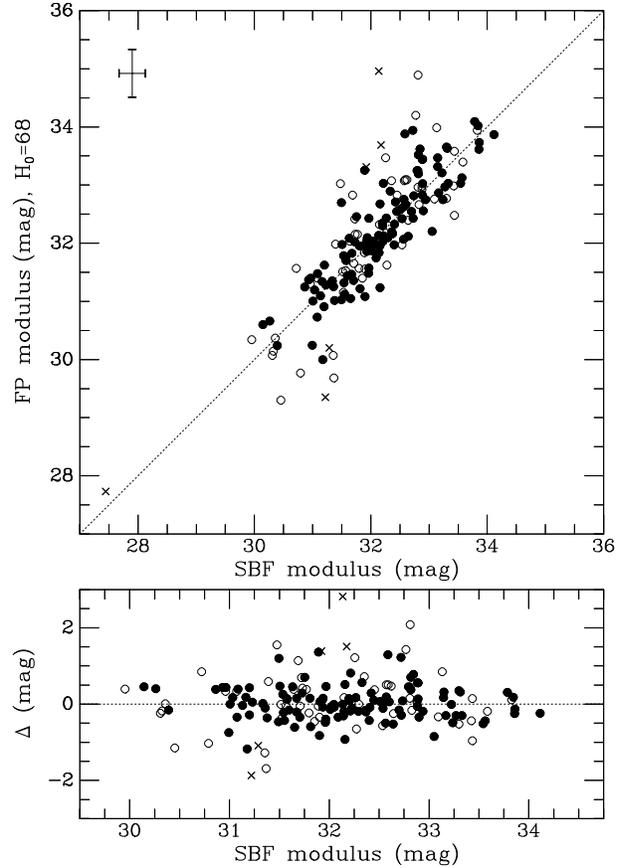}
}\caption{\small
Comparison of the Malmquist-corrected FP and SBF distance moduli
for the 170 galaxies in the cross-matched sample.
The upper panel shows the direct comparison, and the lower
panel shows the residuals (FP modulus minus SBF modulus)
with an expanded horizontal axis.  The FP distances are converted
to Mpc using the best fitting $H_0$ for this comparison.
The median errors are indicated at upper left, but note
that the SBF distances do not all have the same error.
The dotted lines indicate equality, and are not fits to the points.
The different symbols represent ellipticals (filled circles),
S0s (open circles), and the six galaxies excluded from our
$\chi^2$ fits (crosses). Of the six excluded galaxies,
four have velocity dispersions that are either very low
($\sigma < 70$ \kms) or apparently in error (NGC\,3641),
and thus have suspect FP distances; two have very poor
quality SBF observations ($\PD=3.6$) which apparently
bias the distances.  The four nearby, non-excluded S0s
with $\Delta({\rm FP-SBF}) < -1.0$ mag, in order of increasing
$\mM_{\rm SBF}$, are NGC\,3489 (Leo~I?), NGC\,6684,
NGC\,4382 (Virgo), and NGC\,1553 (Dorado).
\label{fig:fpsbfdist}}}
\end{figure}

\begin{figure}
\medskip\vbox{\centering\leavevmode\hbox{
\epsfxsize=7.8cm
\epsffile{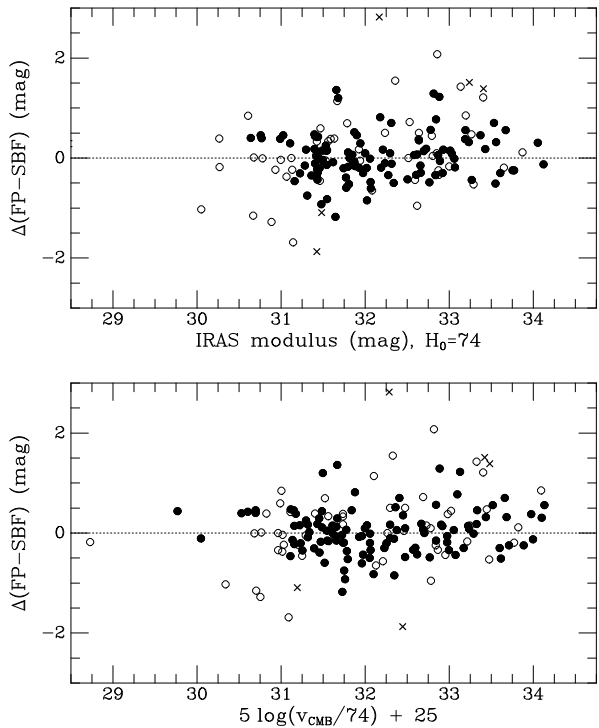}
}\caption{\small
The residuals of the FP-SBF distance comparison are plotted
against \iras-predicted distance modulus (top) and CMB-frame
velocity converted to a distance modulus (bottom).
The value of $\ho=74$ used for the horizontal axis is unimportant, as
different values would simply shift all the points uniformly. 
Symbols are as in Figure~\ref{fig:fpsbfdist}.
\label{fig:fpsbfresids}}}
\end{figure}

Figure~\ref{fig:fpsbfdist} plots the FP distances against the SBF distances.
For the FP distances, we have assumed a
Hubble velocity of 7355~\kms\ (the best-fitting SMAC value) for Coma and the
FP zero point found in \S\ref{ssec:fpdist}.  With these assumptions, we
find that $H_0\approx68$ \kmsM\ minimizes $\chi^2$ for this distance
comparison, and have adopted this value for the figure.  The distances
have been corrected for inhomogeneous Malmquist bias as described in the
previous section; the small covariance introduced by using the same
density field for these corrections is ignored (different errors go
into the corrections). The agreement in Figure~\ref{fig:fpsbfdist}
is generally good, although the scatter
is larger for the 53 S0s (0.70 mag, or 0.65~mag excluding the one
${>\,}2\,$mag outlier) than for the 111 ellipticals (0.45~mag), where we
define the ellipticals as having a morphological $T$-type of $-5$ in the
RC3 (de Vaucouleurs \etal\ 1991).

The larger scatter for the S0s is mainly due to a group of particularly
low-luminosity galaxies with $\mM_{\rm SBF} \approx 31.0$ but $\mM_{\rm
FP} \approx 29.7$.  The S0s cause an apparent trend for the residuals
$\Delta\mM_{\rm FP-SBF}$ to increase with distance.  This is made more
clear in Figure~\ref{fig:fpsbfresids}, which plots the residuals against
the \iras-predicted distances and the CMB-frame velocities.  The trend of
increasing $\Delta\mM_{\rm FP-SBF}$ with $\mM_{\it IRAS}$ is formally
significant at 95\% confidence for the full sample but only 42\%
confidence (i.e., not significant) for just the ellipticals.

Table~\ref{tab:distcomp} gives more information on this and subsequent
distance comparisons, including comparisons for the following
subsamples:  ellipticals, galaxies with $v_{\rm CMB} \le 3000$ \kms,
galaxies with $\viz>1.135$ and intersections of these.  It includes
the number of galaxies for the comparison,
the best-fitting value of $H_0$ and the reduced $\chi^2$ for each comparison,
and the formal slope and reduced $\chi^2$ for a bivariate linear fit
of, for example, FP distance versus SBF distance.
These results are discussed in greater detail in the following section
on systematic effects.  

In calculating the values of $\chi^2$ (and thus the best-fitting $H_0$, etc),
we used the errors described above for the SBF and \iras\ distances, and a
fixed 19.0\% distance uncertainty for the FP, which nominally includes an
intrinsic scatter of 18\% and a 7\% distance error from \xfp\ uncertainties
(Paper~I).  Overall, the FP distance error would have to be
21.7\% to give $\chi^2_\nu\equiv1$ for the FP-SBF comparison, assuming the
SBF errors are accurate, but this neglects the obvious problems with some
S0s.  For just the 111 ellipticals, a distance error of 17.6\%
would give $\chi^2_\nu\equiv1$.

\begin{figure}
\medskip\vbox{\centering\leavevmode\hbox{
\epsfxsize=8.0cm
\epsffile{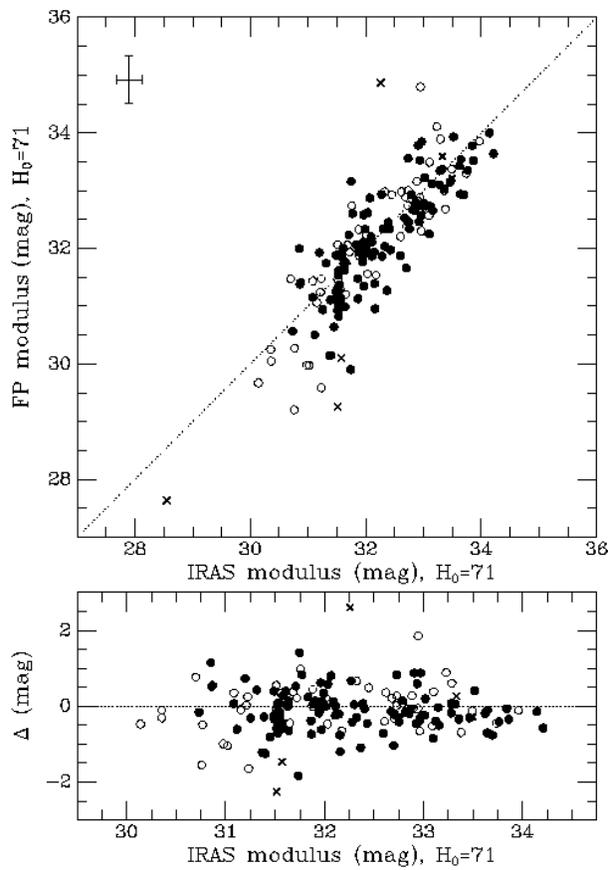}
}\caption{\small
Same as Figure~\ref{fig:fpsbfdist} but for the FP-\iras\ distance comparison.
\label{fig:fpirasdist}}}
\end{figure}

\begin{figure}
\medskip\vbox{\centering\leavevmode\hbox{
\epsfxsize=8.0cm
\epsffile{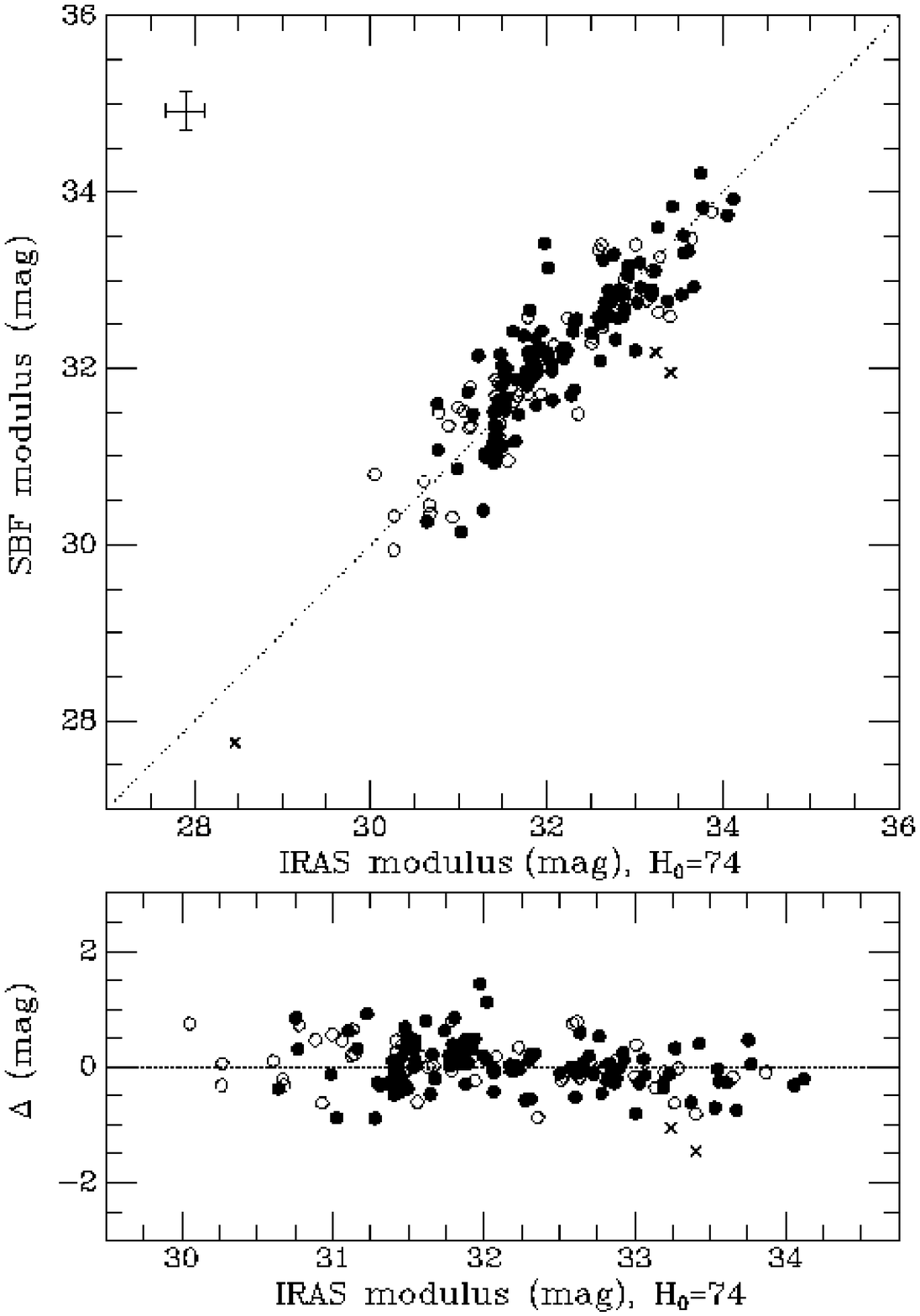}
}\caption{\small
Same as Figure~\ref{fig:fpsbfdist} but for the SBF-\iras\ distance comparison.
\label{fig:sbfirasdist}}}
\end{figure}

Figures~\ref{fig:fpirasdist} and~\ref{fig:sbfirasdist} plot the FP and SBF
distances against the \iras\ distances, and Table~\ref{tab:distcomp} again
summarizes the results.  It would of course be possible to plot all
galaxies with SBF or FP distances against the \iras-predicted distances,
but our intent was to use the \iras\ distances as a tool for diagnosing
the problems and biases in the FP-SBF distance comparisons.  Previous work
(SBF-III; Willick \& Batra 2001) have made comparisons of the \iras\
predictions with the full SBF survey data set.
Figure~\ref{fig:fpirasdist} shows that the same problem with the nearby
S0s occurs in the FP-\iras\ comparison as in the FP-SBF comparison, and
this causes the slope of the FP versus \iras\ distances to be $2\,\sigma$
greater than unity for the full sample.  As one might expect given that a
fixed FP distance error was used for all galaxy types, $\chi^2_\nu$
decreases for the FP-\iras\ comparison when only the ellipticals are
considered. Yet, it unexpectedly increases by 10\% when this is done for the
SBF-\iras\ comparison.  Five of the six galaxies that are more than
2$\,\sigma$ discrepant in this comparison are ellipticals.
It is possible that the peculiar velocities of the ellipticals are more
prone to small-scale non-linear effects and that the smoothing of the
density field underpredicts the local overdensity, or that ellipticals
preferentially occupy regions of non-trivial biasing with respect to the
\iras\ density field.  For such situations, Eq.\,(\ref{eq:sigvar}) would
become inadequate. The poorer FP distances for the S0s would hide this
effect in the FP-\iras\ comparison.
We plan to address this issue more fully in a future paper.

\begin{figure}
\medskip\vbox{\centering\leavevmode\hbox{
\epsfxsize=8.0cm
\epsffile{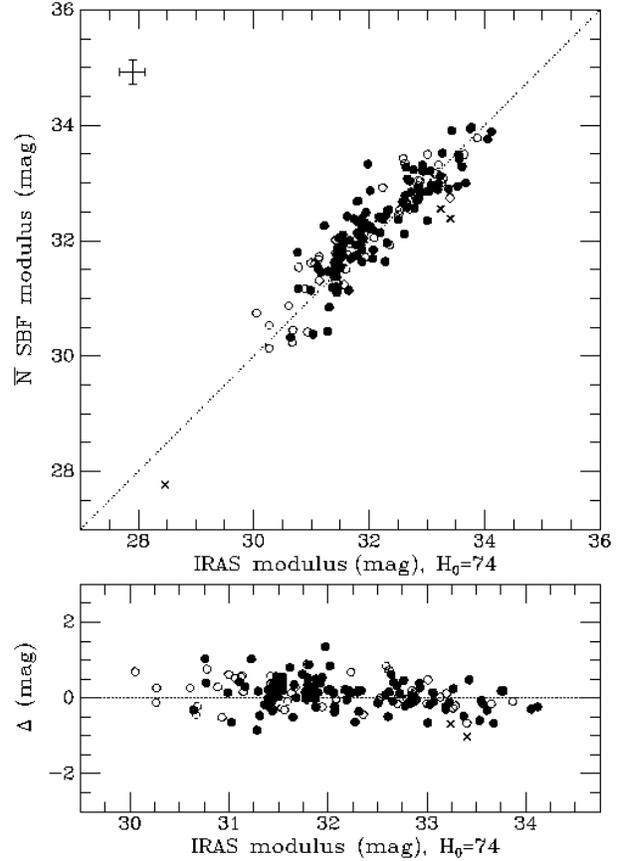}
}\caption{\small
Same as Figure~\ref{fig:fpsbfdist} but for the
\Nbar-calibrated SBF-\iras\ distance comparison.
\label{fig:nbarirasdist}}}
\end{figure}

As noted in \S\ref{ssec:sbfdistances} above, comparisons between the
hybrid \Nbar-calibrated SBF distances and the FP distances would be
misleading and difficult to interpret because of the strong covariance
between \xfp\ and $m_{\rm tot}$.  However, we can compare the \Nbar\,SBF
distances with those predicted from the \iras\ density field, as shown in
Figure~\ref{fig:nbarirasdist}.  Because the intrinsic scatter in the
\mi--$m_{\rm tot}$ relation is not well characterized, and we do not
have an extensive set of homogeneous external $m_{\rm tot}$ values to
gauge our accuracy, we simply adopt the same error for the \Nbar\,SBF
distances as for the \viz-calibrated SBF distances, except we leave out
the contribution from reddening.  This simplifies the intercomparison
of $\chi^2$ from the two analyses.

Surprisingly, the \Nbar\ calibration of SBF actually improves the agreement
with \iras\ over that achieved with the usual \viz\ calibration.  The
scatter in Figure~\ref{fig:nbarirasdist} is 0.36~mag for the S0s and 
0.38 mag for the ellipticals, as compared to 0.40~mag and 0.41~mag,
respectively, in Figure~\ref{fig:sbfirasdist}.  Tables~\ref{tab:distcomp}
shows the improvements in $\chi^2_\nu$.  This reflects how hard it is to
determine \viz\ to sufficient accuracy (including the correction for
Galactic reddening) as well as the decreased sensitivity to \mi\ errors
(Eq.\,\ref{eq:nbarmbar}).

\begin{figure}
\medskip\vbox{\centering\leavevmode\hbox{
\epsfxsize=9.0cm\epsffile{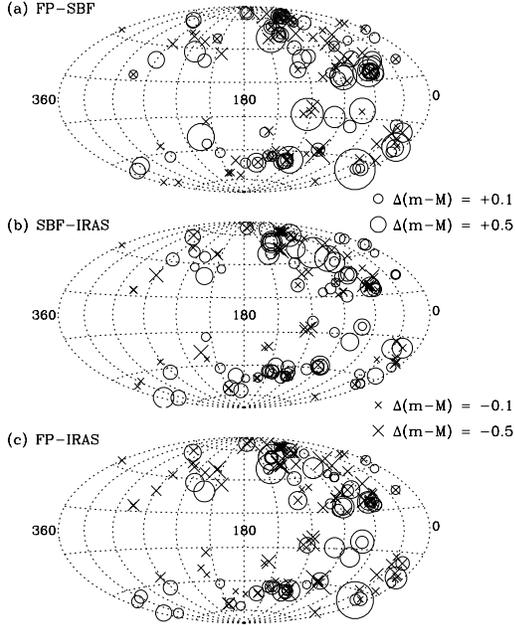}
}\caption{\small
Sky distribution (Aitoff projection) of the distance modulus
residuals in Galactic coordinates for the three
sets of distance comparisons: 
(a) $\mM_{\rm FP} - \mM_{\rm SBF}$ for $H_{0} = 68$ \kmsM,
(b) $\mM_{\rm SBF} - \mM_{\it IRAS}$ for $H_{0} = 74$ \kmsM, and 
(c) $\mM_{\rm FP} - \mM_{\it IRAS}$ .  
Circles show positive residuals and crosses show negative residuals.
Symbol size is proportional to $[0.5+|\Delta(m{-}M)|]$, so that zero
residuals are shown as circles of finite radius.
\label{fig:skyresid}}}
\end{figure}

Figure~\ref{fig:skyresid} shows the sky distribution of the distance
modulus residuals for the three sets of distance comparisons,
where the SBF distances are calibrated using \viz.  For
comparison to the SBF distances, the FP and \iras\ distances are
converted to Mpc using the respective best-fitting values of \ho.
This type of residual plot emphasizes the outliers by
showing them with larger symbols;  the six galaxies excluded from the
$\chi^2$ analyses for the reasons given above are not
shown here.  Positive (circles) and negative (crosses) residuals appear
fairly well mixed in all three panels of Figure~\ref{fig:skyresid},
indicating that there are no major direction-dependent systematic errors
in the three sets of distance estimates.

Overall, the different distance indicators agree quite well and are
consistent within the errors so that the comparisons yield
$\chi^2_\nu\sim1$.  The good agreement allows us to discern possible
systematic problems in the data.  We explore this issue in
greater detail in the following section.

\section{Discussion}

\subsection{The Velocity Field and $\beta_I$}

\begin{figure}
\medskip\vbox{\centering\leavevmode\hbox{
\epsfxsize=8.0cm
\epsffile{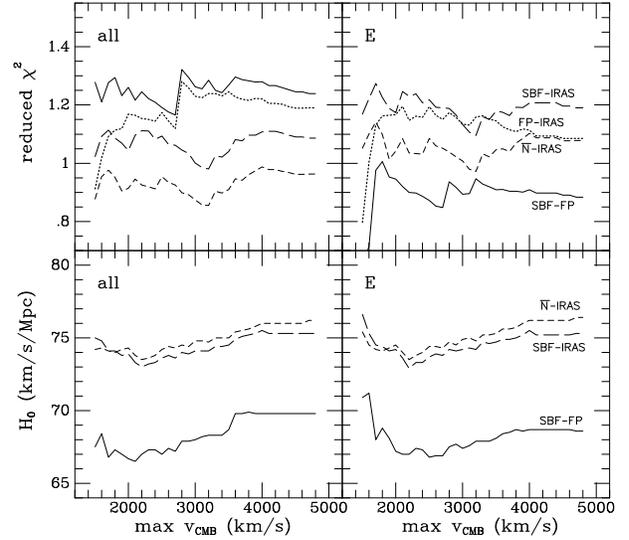}
}\caption{\small
The reduced $\chi^2$ (top panels) and the best-fitting $H_0$ 
values (lower panels) for the various distance-distance
comparisons are shown as a function of the cutoff 
CMB frame velocity for the comparisons. The comparisons
are labelled. Left panels show results from the full sample
regardless of morphological type; right panels show the 
results when the sample is restricted to ellipticals.  
\label{fig:vmax}}}
\end{figure}

Figure~\ref{fig:vmax} shows the dependence of the reduced $\chi^2$ and
best-fitting \ho\ values from the different comparisons of
Table~\ref{tab:distcomp} on the CMB-frame velocity limit.  The
$\chi^2_\nu$ values for the different comparisons are very stable against
the velocity cut.  However, there does appear to be some tendency for
$H_0$ to increase for $v_{\rm CMB} \gta 3000$ \kms, which is significant
at the 1--2\,$\sigma$ level; a possible explanation is discussed in the
following section.  Perhaps the most striking aspect of
Figure~\ref{fig:vmax} is that although there is a $\sim30$\% improvement in
$\chi^2_\nu$ when the SBF-FP comparison is restricted to ellipticals,
the inferred \ho\ changes by less than 2\%.

\begin{figure}
\medskip\vbox{\centering\leavevmode\hbox{ \epsfxsize=8.0cm
\epsffile{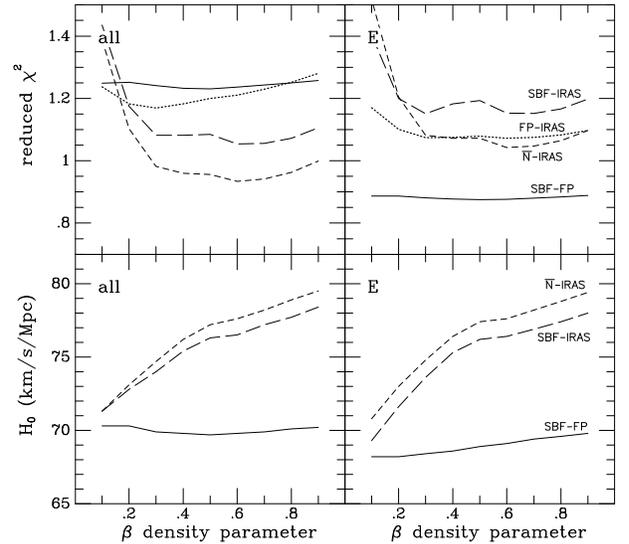} }\caption{\small The reduced $\chi^2$ (top panels)
and the best-fitting $H_0$ values (lower panels) for the various
distance-distance comparisons are shown as a function of the \iras\
$\beta$ parameter used in the galaxy density field reconstruction and
peculiar velocity predictions.  Left panels show the results for all
morphological types; right panels show the results when the sample
is restricted to ellipticals.  The SBF-FP comparisons are
insensitive to $\beta$ because the density field is only used for the
inhomogeneous Malmquist corrections to these distances, but it is the
basis of the \iras-predicted distances.  No velocity restriction was
applied to the galaxies for these comparisons.
\label{fig:beta}}}
\end{figure}

\begin{figure}
\medskip\vbox{\centering\leavevmode\hbox{
\epsfxsize=8.0cm
\epsffile{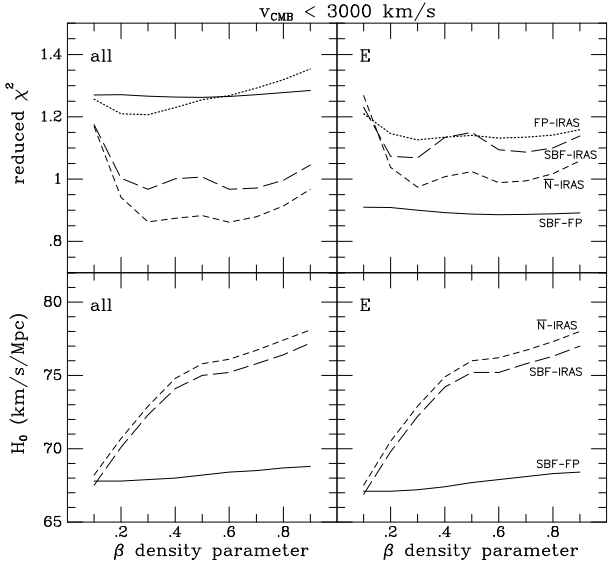}
}\caption{\small
Same as Figure~\ref{fig:beta} except the comparisons have
been restricted to galaxies with CMB frame velocities
$v_{\rm CMB}{\,<\;}3000\,$\kms.
\label{fig:beta3000}}}
\end{figure}

Figure~\ref{fig:beta} plots $\chi^2_\nu$ and $H_0$ from the same
comparisons against the value of $\beta_I$ used for the velocity field
reconstruction and the density term in the Malmquist corrections.
Figure~\ref{fig:beta3000} is the same, but restricted to galaxies with
$v_{\rm CMB}<3000$ \kms.  The SBF-FP comparison shows reassuringly little
dependence on $\beta_I$.  The lower panels of Figures~\ref{fig:beta}
and~\ref{fig:beta3000} indicate that \ho\ from the SBF-\iras\ comparison
can be brought into agreement with that from the SBF-FP comparison for
$\beta_I\approx0.1$.  However, the SBF-\iras\ comparison strongly rejects
values of $\beta_I<0.3$, as do most other analyses of this sort, which
typically give $\beta_I$ in the 0.4--0.9 range
(e.g., da\,Costa \etal\ 1998; Willick \& Strauss 1998; Riess \etal\ 1997;
Sigad \etal\ 1998; Hudson \etal\ 1999).  Thus, there is little prospect
for reducing the disagreement on \ho\ to less than~9\%.

There appears to be little constraint on larger values of $\beta_I$ from
these data {\it if \ho\ is allowed to increase in tandem}, and the constraints
are weakened when restricted to $v_{\rm CMB}<3000$ \kms.  However, when
high-quality data at large distance from \hst\ are introduced, the
covariance between \ho\ and $\beta_I$ is reduced and the constraints are
strengthened.  SBF-III concluded that the $1\,\sigma$ uncertainty on \ho\
from $\beta_I$ itself, and thus the \iras-velocity tie, was only about
1.5\%.  However, this does not include possible systematic errors in 
the \iras\ redshift sample used in deriving the density field; 
see~\S\ref{ssec:whichH0} below.

\subsection{Selection Bias in the SBF Survey?}
\label{ssec:sbfselect}

SBF-II discussed a possible bias in the SBF survey data, such that the
poorest quality data preferentially produced systematically low distance
estimates.  The main data-quality indicator they used was \PD, which is
the product of the seeing full-width at half-maximum in arcseconds and the
CMB frame velocity in units of 1000 \kms\ (so $[\PD]^2$ goes as the metric
area within a resolution element).  Occasionally, repeat observations with
high values of $\PD\sim3$ gave erroneously low distances.  Because of this
perceived `PD bias,' SBF-II used only data with $\PD<2.7$ for their
analysis, and SBF-III adopted the same \PD-cutoff.  SBF-IV included
results for observations with $2.7<\PD<3.0$ in a separate table for
`uncertain SBF data.'

A bias in the poorest quality data is reasonable to expect if the SBF
survey is viewed as \mbar-limited.  Near the \mbar\ limit of the
survey, if the SBF signal is not detected (inferred distance too big) the
galaxy does not make it into the catalogue, but if the SBF {\em is} detected
(inferred distance smaller), it does make it in.  Galaxies with the
largest true distances in the survey will have measured distances that are
too low, because their true \mbar\ values will be fainter than the survey
limit, and observational error has scattered them into the survey catalogue
(these might be called `spurious SBF detections').
This selection bias at the faint limit of the survey is distinct from the
Malmquist bias which occurs at all distances (although both biases are
rooted in normal observational error and are minimized when these errors
are small).

Because the exposure times and seeing conditions of the SBF observations
were tailored to the expected distances of the galaxies, the survey data
are not actually \mbar-limited, but we should still expect this sort of
selection bias in marginal SBF detections.  The only way to avoid it would
be to volume-limit the sample with a conservative cutoff, or to impose a
volume limit as a function of the seeing in the data image.  Effectively,
this is what the $\PD<2.7$ cutoff of SBF-II and SBF-III did: impose a
volume limit of 2700 \kms\ for 1\arcsec\ seeing and allow this limit to
scale with seeing so that it was about Virgo distance for 2\arcsec\
images and $\sim4000$ \kms\ for the best (0\farcs65--0\farcs7) seeing.

As seen in Figures~\ref{fig:fpsbfresids} and \ref{fig:sbfirasdist}, there
may be some evidence for a bias in the SBF distances at the largest `true'
distances (gauged by CMB velocity or \iras-predicted distance).  This
effect is also seen in the increasing inferred \ho\ for increasing
$v_{\rm CMB}$ cutoffs in Figure~\ref{fig:vmax}.  However, the overall
slope (no distance or velocity limit) of a bivariate fit to the SBF-\iras\
distance modulus comparison is unity, $1.00\pm0.03$ (Table~\ref{tab:distcomp}).
The slope for a bivariate of the full SBF-FP comparison is $3\sigma$ from
unity, but this is largely driven by the nearest S0s; when the S0s are
excluded, the slope is within $1\sigma$ of unity.

\begin{figure}
\medskip\vbox{\centering\leavevmode\hbox{
\epsfxsize=7.8cm
\epsffile{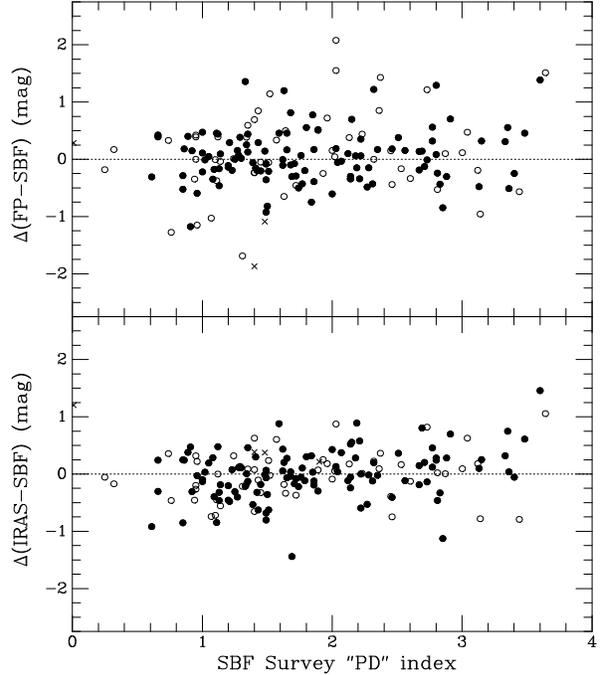}
}\caption{\small
The difference between the FP and SBF distance moduli (top)
and \iras-predicted and SBF distance moduli (bottom) are
plotted against the SBF data quality index \PD\ from SBF-II.
The symbol types are as in Figure~\ref{fig:fpsbfdist}, except
there is no need to distinguish the two (apparently biased)
$\PD>3.5$ galaxies with crosses here.
\label{fig:pd_dresid}}}
\end{figure}

\begin{figure}
\medskip\vbox{\centering\leavevmode\hbox{
\epsfxsize=7.6cm
\epsffile{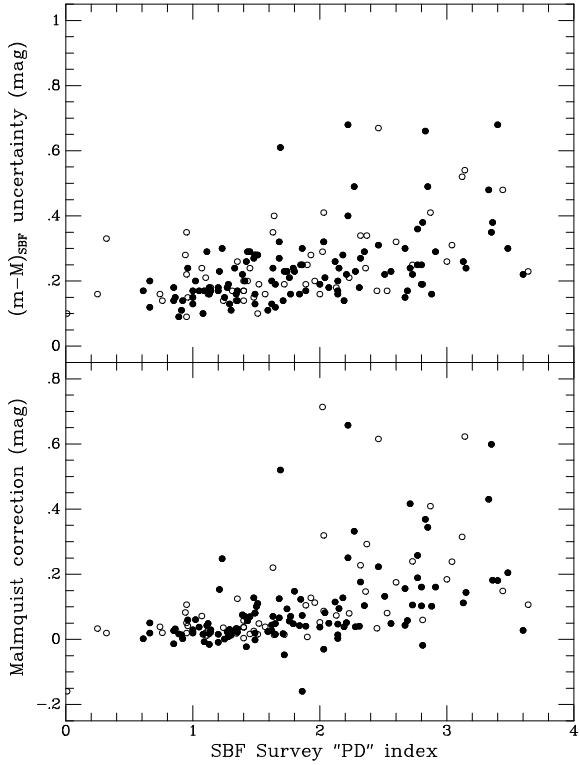}
}\caption{\small
The SBF survey distance uncertainties (top) and the inhomogeneous
Malmquist bias corrections determined in Section~\ref{ssec:malmquist} are
plotted as a function of the \PD\ data quality index from SBF-II.  Open
symbols are for ellipticals and filled symbols are for S0s.  The Malmquist
corrections increase with \PD\ because the poorer quality data have
larger errors on average.
\label{fig:pderror}}}
\end{figure}

Figure~\ref{fig:pd_dresid} shows the residuals of the FP-SBF and
SBF-\iras\ comparisons as function of \PD.  As plotted, the residuals will
be preferentially positive at large \PD\ if there is a significant
selection bias.  It appears that the $\PD>2.7$ selection bias in the SBF
survey data is not as bad as feared by SBF-II.  This may be because SBF-II
did not correct for Malmquist bias, which, although distinct from the
selection bias, increases with \PD\ because the observational errors do.
Figure~\ref{fig:pderror} illustrates this point.  Consequently, any bias
parameterized by \PD\ will be a combination of Malmquist and selection
biases and will be reduced following Malmquist correction.  That said,
there remains a suggestion of SBF selection bias in the SBF-\iras\
comparison of Figure~\ref{fig:pd_dresid}: 11 of 14 galaxies, including 7
of 8 ellipticals, with $\PD>3.0$ have positive residuals, and the two with
$\PD\approx3.6$ may well be `spurious SBF detections.'  However, it is
important to note that the data with $\PD>3.0$ were judged as too poor to
be included even in the `uncertain data' table of SBF-IV.  We conclude
that, following the inhomogeneous Malmquist corrections applied here, SBF
data with $\PD<3$ show very little evidence of selection bias.

\subsection{Stellar Population Effects} 
\label{ssec:SPeffects}

The SBF method makes an explicit correction for stellar population in
terms of \viz\ colour. The FP method, as applied here and in most other
works, assumes that all early-type galaxies {of a given mass} have
essentially the same stellar population, i.e., that the mass-to-light 
ratio scales in a consistent way with mass.  Variations in the stellar
content, such as from recent star formation, will
cause deviations from the FP at some level.  Here, we examine the
possibility of trends in the distance residuals with \mgii\ index
and \viz\ colour.

\begin{figure}
\medskip\vbox{\centering\leavevmode\hbox{
\epsfxsize=8.0cm
\epsffile{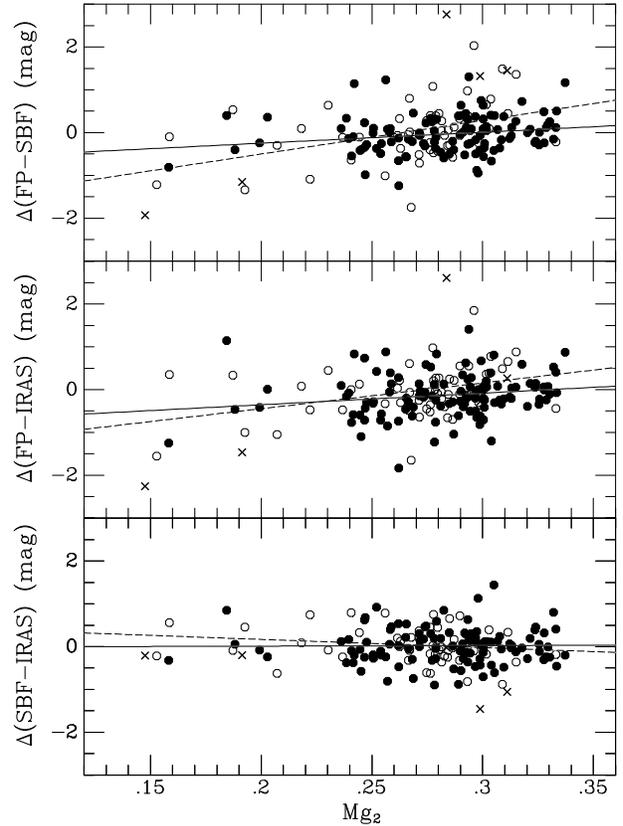}
}\caption{\small
The differences between the FP and SBF distance moduli (top),
FP and \iras-predicted moduli (middle), and 
SBF and \iras-predicted moduli (bottom) are plotted against \mgii.
The symbol types are as in Figure~\ref{fig:fpsbfdist}. 
Solid and dashed lines are simple least-squares fits to the elliptical
and S0 subsamples, respectively (not to the combined sample).
\label{fig:dFPres_mg}}}
\end{figure}

Figure~\ref{fig:dFPres_mg} plots the residuals of the FP-SBF, FP-\iras,
and SBF-\iras\
distance comparisons against \mgii.  Overall, the trend of increasing
$\Delta\mM_{\rm FP-SBF}$ with \mgii\ is significant at the 3.7\,$\sigma$
level, or 2.3\,$\sigma$ for the ellipticals and 3.6\,$\sigma$ for the S0s.
The trend of increasing $\Delta\mM_{\rm FP-{\it IRAS}}$ with \mgii\ is
significant at 2.2\,$\sigma$, or 2.0\,$\sigma$ for the ellipticals and
2.8\,$\sigma$ for the S0s.  These trends appear to be real, 
but mild and not subject to simple interpretation.
There is no significant trend in the SBF-\iras\ distance 
residuals with \mgii.

\begin{figure}
\medskip\vbox{\centering\leavevmode\hbox{
\epsfxsize=8.0cm
\epsffile{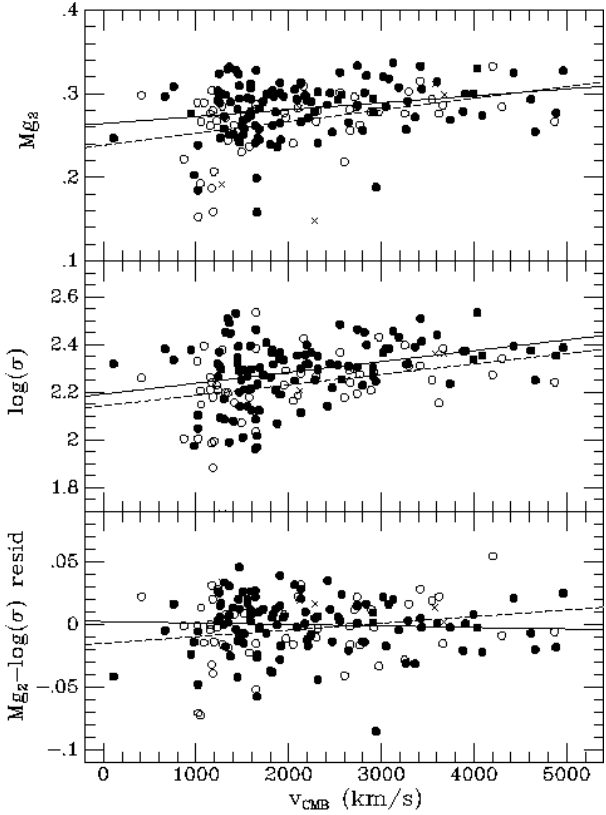}
}\caption{\small
\mgii\ (top), $\log\sigma$ (middle), and residuals with respect to
the mean \mgii--$\sigma$ relation from Paper~I
(bottom) are plotted against CMB-frame velocity.  The residuals
in the bottom panel are positive (negative) when the observed \mgii\ 
is greater (lower) than the predicted value from the mean relation
and the observed $\log\sigma$.
Symbol types are as in Figure~\ref{fig:fpsbfdist}. Solid and dashed
lines are simple least-squares fits to the elliptical and S0 
subsamples, respectively (not to the combined sample).
\label{fig:mgsig_vcmb}}}
\end{figure}

\begin{figure}
\medskip\vbox{\centering\leavevmode\hbox{
\epsfxsize=8.0cm
\epsffile{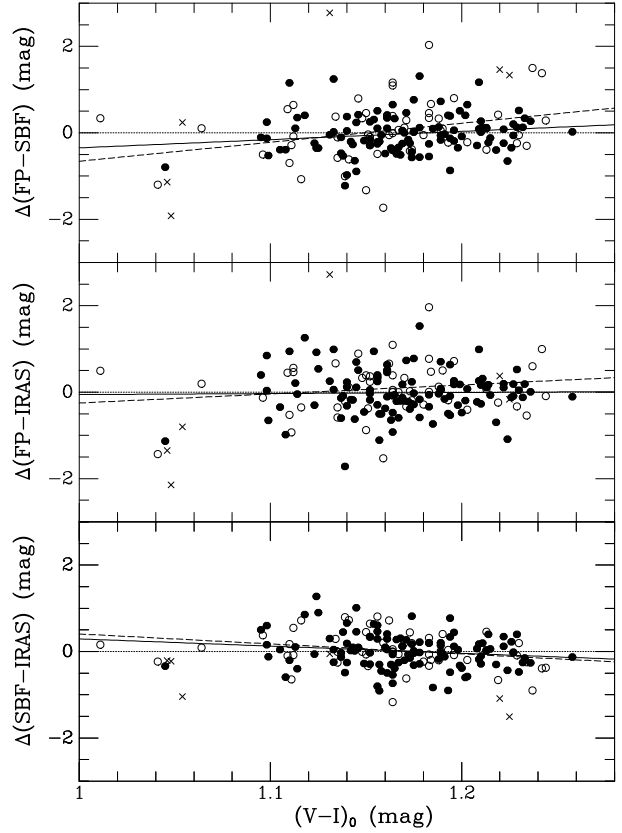}
}\caption{\small
Distance residuals for the FP-SBF, FP-\iras, and SBF-\iras\
comparisons are plotted against \viz\ colour.
Symbol types are as in Figure~\ref{fig:fpsbfdist}. Solid and dashed
lines are simple least-squares fits to the elliptical and S0 
subsamples, respectively (not to the combined sample).
\label{fig:vi_dresid}}}
\end{figure}

Because \mgii\ correlates so strongly with $\log\sigma$, these
trends may simply reflect sample selection effects,
and not indicate a problem with the individual distances.  For instance, the
slope $\alpha{\,=\,}1.22$ determined from the Coma cluster was used for
the FP distances even though it does not give the formal (naive) `best fit'
to the FP of our more inhomogeneous sample.  The shallower slopes of
$\alpha\approx1.0$--1.1 (e.g., Figure~\ref{fig:irasFPcal}) found by
calibrating the FP for the current sample using SBF and \iras\ distances
would decrease the significance of the trends with \mgii\ to 
$\sim\,$2.8\,$\sigma$ for FP-SBF and $\sim\,$1.8\,$\sigma$ for FP-\iras,
but we avoided these shallower slopes because they are determined
from an inherently biased sample.  

Figure~\ref{fig:mgsig_vcmb} illustrates the situation, showing that the
luminosity-based sample selection causes the mean $\log\sigma$ and \mgii\
to increase with distance.  What we should test for then is some
systematic trend of the \mgii--$\log\sigma$ {\it residuals} with distance,
to see if stellar population effects are causing a distance-dependent bias
in the FP distance estimates. We assume the \mgii-$\log\sigma$ relation
found in Paper~I.  As the bottom panel of
Figure~\ref{fig:mgsig_vcmb} indicates, we find no significant overall
trend of the residuals with distance.  The slope of the residuals with
$v_{\rm CMB}$ is within $\sim\,$0.6$\,\sigma$ for the full sample and the
elliptical subsample (for which the best-fitting slope actually goes in the
opposite sense).  

For the S0 subsample (again, including the $T{\,=\,}-4$ extended cD galaxies),
the slope deviates from zero by 1.8\,$\sigma$.  However, this is
mainly the result of a few outliers.  The largest positive residual is
NGC\,3311 at $v_{\rm}=4200$ \kms.  In fact, this galaxy well illustrates
the power of the fundamental plane.  NGC\,3311 is a massive, extended, low
central surface brightness cD in Hydra; its velocity dispersions is
only $\sigma=190$ \kms, which is consistent with the low \sbe, but causes
a deviation from the \mgii--$\log\sigma$ relation because the central
dispersion is not a good indicator of mass for this galaxy.  
We also note that NGC\,3311 has some dust in its centre, although
this appears confined to radii $\lta5\arcsec$, much smaller than
the effective radius $R_e\approx100\arcsec$.  
In any case, the dust would not affect the galaxy's position in
Figure~\ref{fig:mgsig_vcmb}, and does not cause any significant discrepancy
between its SBF and FP distances, which agree to within 0.5\,$\sigma$.

On the other hand, the two most 
significant negative residuals at 1000~\kms\ (NGC\,3489, NGC\,4382)
are nearby S0s 
having FP distances significantly smaller than the SBF or \iras\ distances;
thus, these galaxies do
deviate from the FP, and they deviate because their
velocity dispersions are too low for their luminosities (causing
underestimated FP distances).  
The excess luminosity could result from recent star formation, which
would decrease their \mgii\ below the value expected at a given 
$\log\sigma$, consistent with Figure~\ref{fig:mgsig_vcmb}.
We conclude that the few nearby S0s which cause the weak trend in
\mgii-$\log\sigma$ with redshift deviate from the general relation,
at least in part, because of their stellar populations,
although anomalous virial properties for low-mass S0s could contribute.
In addition, errors in their $\log\sigma$ aperture corrections 
may also play a role (see \S\ref{ssec:aperture}).

For comparison to the \mgii\ plots, Figure~\ref{fig:vi_dresid} plots
the distance residuals against \viz\ colour.  The only trend that is
significant at more than 2$\sigma$ here is the increase of 
$\Delta\mM_{\rm FP-SBF}$ with \viz, which is 3\,$\sigma$ significant
for the full sample and S0 subsample, 2\,$\sigma$ significant
for the elliptical subsample.  There are two reasons for this trend:
first, the nearby galaxies with the underestimated FP distances 
tend to be blue; second, and similar to the above case for the FP and
the shallower best-fitting slope against $\log\sigma$, this sample
would call for a naive best-fitting slope for \Mi\ versus \viz\ that is
somewhat shallower than the 4.5 found from the
maximum likelihood analysis of galaxies in groups by SBF-I.
In any case, the trend of the SBF-\iras\ residuals with \viz\ is
not significant at the 2\,$\sigma$ level.

It is noteworthy that, if anything, the bluer galaxies in
Figure~\ref{fig:vi_dresid} tend to have underestimated FP distances;
this would not be the case if the negative \xfp\ residuals found
in the comparison to the Faber \etal\ (1989) data by Paper~I had
resulted from a bias in our fits to the SBF survey photometric data.
If that were the case, our FP distances for these galaxies would
be biased high, but there is no evidence of that here.
The reduced $\chi^2$ values for the distance comparisons do
improve when the bluer galaxies are excluded (Table~\ref{tab:distcomp}),
but again, this is because the nearby galaxies with
underestimated FP distances are mainly blue S0s.

\subsection{Galactic Extinction}

\begin{figure}
\medskip\vbox{\centering\leavevmode\hbox{
\epsfxsize=8.0cm
\epsffile{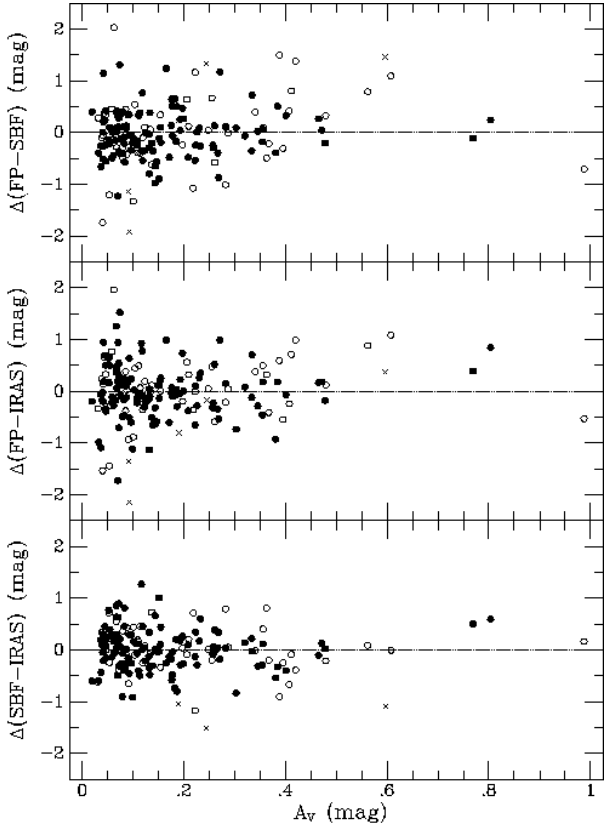}
}\caption{\small
Distance residuals for the FP-SBF, FP-\iras, and
SBF-\iras\ comparisons are plotted against $V$-band
extinction from SFD.
Symbols are as in Figure~\ref{fig:fpsbfdist}.
\label{fig:ext_dresid}}}
\end{figure}

Figure~\ref{fig:ext_dresid} plots distance residuals against $A_V$
extinction from SFD.  As noted in \S\ref{sec:distances}, the SBF distance
will be overestimated, while the FP distance underestimated, if the
extinction is overestimated, and vice versa.  The distances predicted by
the \iras\ flux-limited survey should be essentially independent of SFD
extinction errors, or at least related in a nontrivial way.  While it is
intriguing that the S0s with $A_V \gta0.4\,$mag tend to have larger FP than
SBF or \iras\ distances, if this were an effect of extinction
errors, these galaxies would all have to have a true $A_V$ very close to
zero, which is unlikely.  In short, we find no evidence for a dependence
of the residuals on $A_V$, but the distance errors are too large for a
very significant test.  Better constraints on extinction errors can be
found from the residuals of the \mgii-colour relation
(e.g., SFD; Hudson 1999; Paper~I).

\subsection{Aperture Effects}
\label{ssec:aperture}

The SMAC survey estimated aperture corrections for the
velocity dispersion data using
\begin{equation}
(\log\sigma)_{\rm cor} \;=\; (\log\sigma)_{\rm obs} \,+\,
 0.04 \log\left(r_{\rm ap} \over r_{\rm norm}\right)\,,
\label{eq:apcor}
\end{equation}
where the normalization radius $r_{\rm norm}$ is the angular size
corresponding to 0.6 \hkpc\ and $r_{\rm ap}$ is the radius of a circular
aperture equivalent to the spectroscopic extraction aperture.
The logarithmic slope of 0.04 is taken from Jorgensen \etal\
(1995), who used an extensive set of empirical models to derive mean
aperture corrections.  These authors quoted a FWHM for the logarithmic
slope distribution of 0.06 (0.01--0.07) and found a weak correlation
of the slope with $\log\sigma$ such that a mean of $\sim\,0.03$ was more
appropriate for low dispersion galaxies and $\sim\,0.05$ was more
appropriate for high dispersion ones. 

Our FP measurements are tied to the Coma cluster, which is $\sim\,$8 times
more distant than the closest galaxies in the present survey, yet the
galaxies in our sample had their dispersions measured in apertures not
much larger than used for Coma. Aperture corrections therefore have the
potential for being a significant source of systematic error, and one
which we have not yet considered.  We wish to address the following two
questions.  First, could aperture effects be the cause of the apparently
underestimated distances for some of the nearby, low-dispersion galaxies?
This includes the S0s listed in the caption to Figure~\ref{fig:fpsbfdist}
and the two galaxies at a similar distance which were excluded (and
shown as crosses in the distance comparison figures) specifically because
of their low dispersions.
Second, how large is the systematic uncertainty in the tie to the Coma
cluster, and therefore the value of $H_0$, resulting from the aperture
corrections?

For the nearby discordant galaxies, the aperture correction given by
Eq.\,(\ref{eq:apcor}) for $r_{ap}\approx2\arcsec$ (see SMAC-III) amounts to
about 0.035 dex in $\log\sigma$, or about 10\% in distance.  The assumed
logarithmic slope for $\log\sigma$ needs to be decreased for the nearby
low-dispersion galaxies in order to increase their distances, and this is
qualitatively consistent with the assertion by Jorgensen \etal\ (1995)
that these galaxies have shallower $\log\sigma$ profiles.  However, even
if we took a slope of 0.0, indicating no gradient, their
distances would be increased by only 0.2~mag, not the $\sim1$~mag required
to bring these galaxies into agreement with the SBF and \iras\ distances
in Figures~\ref{fig:fpsbfdist} and \ref{fig:fpirasdist}.  Therefore, while
inappropriate aperture corrections may contribute to the discrepancy for
these galaxies, the full explanation likely
involves stellar population effects as well (\S\ref{ssec:SPeffects}).

With regard to $H_0$, we note that the median
redshift of the galaxies in the sample is about 1800 \kms, implying
a median aperture correction of 0.021 dex from Eq.\,(\ref{eq:apcor}),
or a median correction to the distances of 6\%.  If we allow that the
mean slope of the $\log\sigma$ gradients for these galaxies may differ 
by $\pm0.015$ from the 0.040 used in the SMAC catalogue, which seems
reasonable given the range found by Jorgensen \etal\ (1995), as
well as the scatter in $\log\sigma$ gradients observed by
Franx \etal\ (1989), then the derived distance
scale may change by 2.2\%, or about $\pm1.5$ \kmsM\ in \ho.  We use
this in the following section in estimating the systematic 
uncertainty in~\ho.

\subsection{Systematic Effects on $H_0$}
\label{ssec:whichH0}

The value of $H_0$ found by tying the Cepheid-calibrated SBF distances to
the Hubble flow via the FP is $68.0\pm1.3$ from the ellipticals within
3000~\kms\ and all types within 3000~\kms, beyond which point $H_0$ shows
an increase that may reflect selection biases.  The uncertainty in the
Hubble-flow tie via the FP is about 3.5\%, including a $\sim\,$200 \kms\
distance uncertainty for the peculiar velocity of the Coma cluster and
the uncertainty due to aperture effects as described
in the previous section.  With an additional allowance of 2.5\%
from half the range of $H_0$ values from the various FP-SBF comparisons,
we conclude $H_0=68\pm3.2$ \kmsM\ from the SBF-FP comparison.

For the SBF-\iras\ comparison, we adopt $H_0=74\pm1.6$, consistent with
Table~\ref{tab:distcomp} (especially within 3000 \kms), but based also on
the results and $\beta_I$ uncertainty from SBF-III, which considered a
total of 280 galaxies, including several observed with \hst, and made a
much more stringent $\PD<2.7$ quality cut to avoid potential selection
bias.  Thus, the \iras\ and FP ties of SBF to the far-field Hubble flow
give \ho\ values differing by $\sim\,$1.7$\,\sigma$.
Another sympton of the problem, and one which does not involve SBF distances,
is the large Hubble distance of $\sim\,$8000 \kms\ for the Coma cluster
given by the tie between the `\iras-calibrated FP' of 
Figure~\ref{fig:irasFPcal} and the observed Coma FP zero point in
Figure~\ref{fig:ComaFP}.  Note that
the quoted errors on $H_0$ include only the uncertainties on ties between
the relevant methods.

These results are reminiscent of those of Willick \& Batra (2001),
who found that tying the Cepheid distances to the Hubble flow via the
\iras\ peculiar velocity predictions (which they called `Method~2') gave
an \ho\ $\sim\,$10\% greater than tying to the Hubble flow using the
distance ladder of secondary indicators (`Method~1').  The latter,
Method~1, approach was used by the \ho\ Key Project group (e.g., Ferrarese
\etal\ 2000; Gibson \etal\ 2000; Sakai \etal\ 2000; Freedman \etal\ 2001).
(The other half of the $\sim\,$20\% difference between the Key Project and
Willick \& Batra \ho\ values resulted from a systematic difference of
nearly 0.2~mag in their Cepheid distance scales.)

The discrepancy seen here and by Willick \& Batra could be explained if
the \iras\ redshift survey overestimates the density field locally, so
that the local corrected velocities are systematically too high, yielding
a higher \ho.  This would result from a problem in the redshift survey
completeness estimates as a function of distance.  If this could be
demonstrated, then we should prefer $H_0\approx68$ from the FP-SBF
comparison.  However, $H_0 = 74$ is only 1.9$\,\sigma$ from the best-fitting
value from the SBF-FP tie, and could be closer to 1$\,\sigma$ if we had
used slightly larger estimates of the systematic uncertainties.  At this
point, the safest bet might be a straight average of the two results,
$H_0\approx71\pm4$ \kmsM, where the 1-$\sigma$ uncertainty is estimated
from the difference over $\sqrt{2}$.
This \ho\ is consistent with most other recent determinations
from extragalactic distance indicators tied to the same Cepheid scale.

If we update the SBF zero point by $+$0.06 mag according to the revised
Key Project Cepheid distances from Freedman \etal\ (2001), then the Hubble
constants from the FP-SBF and \iras-SBF comparisons become 70 and 76
\kmsM, respectively.  We estimate the systematic error from the 5\%
uncertainty in the tie between SBF and Cepheids (SBF-II), the $\sim\,$10\%
systematic uncertainty on the Cepheid distances (see for example Ajhar
\etal\ 2001), and the additional $\sim\,$10\% uncertainty in the LMC
distance modulus (e.g., Walker 1999; Gibson 2000).  Our final value
for \ho\ from early-type galaxies calibrated against Cepheids is then
$H_0=73\pm4\pm11$~\kmsM.

We note that Freedman \etal\ (2001), based on the work of Kelson
\etal\ (2000a), reported $H_0=82\pm6\pm9$ \kmsM\ from the FP.  However, they
relied upon group association (the Cepheid distances to the Leo group and
Fornax and Virgo clusters) for their distance calibration.  In addition,
their calibrating and distant cluster samples (the latter being taken from
Jorgensen \etal\ 1996) were observed in different bands, came from
independent data sources, and had no photometric or spectroscopic overlap.
Thus, while Kelson \etal\ provided detailed estimates of the systematic
and random uncertainties, the true level of homogeneity between their
calibrating and program samples was unknown.  We believe our results for
\ho\ are more robust for this reason.


\section{Summary and Conclusions}

We have used the \xfp\ photometric parameters derived for the SBF survey
galaxies in Paper~I and the homogenized velocity dispersions from SMAC-III
to obtain FP distances for over half of the SBF survey galaxies.  We
corrected the FP and SBF distances for inhomogeneous Malmquist bias and
intercompared the FP, \viz-calibrated SBF, and \iras-predicted
distances. The distance agreement was good overall, with reduced $\chi^2$
values near unity for most of the comparisons.  However, the inclusion
of S0s significantly increases the scatter and $\chi^2_\nu$ for the
FP-SBF and FP-\iras\ comparisons.  While the mean error per FP measurement
is found to be 21.7\% overall, it is reduced to 17.6\% when just ellipticals
are considered.   This is mainly due to the several nearby,
low-dispersion, blue galaxies whose FP distances are anomalously low.
For the SBF-\iras\ comparison, the S0s actually decrease the scatter.
While the reason for this is unclear, it may be that the smoothed
\iras\ velocity field model is simply inadequate for many ellipticals
and underpredicts the local velocity field noise, so that the
predictions for the S0s are more accurate overall.  

We also compared the \iras\ distances to SBF distances calibrated by the
\Nbar\ fluctuation count parameter introduced by SBF-IV.  Because
\Nbar\ correlates strongly with the FP, \Nbar-based distances can be
considered a hybrid SBF-FP distance indicator; it is therefore not
meaningful to compare \Nbar\,SBF and FP distances.  We find the
\Nbar\,SBF distances exhibit less scatter in the comparison to the \iras\
distances.  Taken at face value, the improvement in $\chi^2$ indicates
that the \Nbar\,SBF distances are roughly 25\% more accurate than
\viz-calibrated SBF distances.  The better accuracy is due to the
decreased sensitivity to photometric errors, \mbar\ measurement errors,
and Galactic extinction.

In addition to the morphological effects, several other sources of
potential systematic error were considered.  These included the density
parameter $\beta_I$ used for the velocity reconstruction and the Malmquist
corrections, selection biases in the SBF survey with distance, Galactic
extinction, and stellar population and aperture effects.  We found that
low values of $\beta_I\lta0.1$ can bring the FP and \iras\ ties to the
Hubble flow into close agreement, but are excluded for other reasons.  We
find some evidence of selection bias near the limit of the SBF survey,
seen in weak increases in \ho\ and in the reduced
$\chi^2$ for the SBF-\iras\ comparison beyond about 3500 \kms.  However,
this bias should not affect the published SBF survey distances or
analyses, as these used fairly stringent quality cuts specifically to
avoid this problem, whereas we have made essentially no quality cuts so
that we could test for such a bias.  

The nearby, blue, mainly S0 galaxies with underestimated FP distances
are apparently affected by young stellar populations,
although possibly coupled with irregular virial properties and/or
inappropriate aperture corrections.  However, apart from these few
nearby S0s, stellar population effects do not appear to significantly
bias the FP distances. 
We also find no evidence that the Galactic extinction estimates 
contribute significant systematic distance errors.

The comparison between the Cepheid-calibrated SBF distances and the far-field
Hubble flow calibrated FP distances yields $H_0=68\pm3$ \kmsM,
including statistical and systematic uncertainties in the SBF-FP tie and
the tie of the FP distances to the Hubble flow.  The SBF-\iras\ comparison
yields $H_0 = 74\pm2$ \kmsM; so, formally there is a $\sim\,$1.7$\,\sigma$
difference between \ho\ given by the two separate ties.  However, additional
sources of systematic error, such as larger uncertainties in the FP
aperture corrections, or possible completeness errors in the \iras\ 
density field used for the velocity reconstruction, would further 
reduce the marginal significance of this difference.  Therefore, we
have recommended a straight average of the above two \ho\ values.
We have noted that the SBF distance zero point is from SBF-II, which
uses the Key Project Cepheid distances tabulated by Ferrarese \etal\ (2000).
Had we used a zero point based on the revised Cepheid tabulation
by Freedman \etal\ (2001), the \ho\ values would be 2.8\% greater.
Therefore, the final average would be $H_0=73\pm4$ \kmsM\ (internal error).
Additionally, there is a systematic error of about 15\% in \ho\ from
the zero point of the Cepheid scale and the tie between SBF and Cepheids.

Future space-based SBF observations of ellipticals in a significant number
of galaxy clusters should allow for further progress on the early-type
galaxy distance scale.  This will enable direct comparison of high-quality
SBF distances to inverse FP measurements of cluster distances, each based
on ten or more galaxies. Such a comparison would be much less sensitive to
Malmquist, selection, and aperture biases.  In addition, further \hst\ SBF
distances in the direction of the Hydra-Centaurus supercluster are greatly
needed.  The ground-based SBF survey do not sample this region well, and,
as we have discussed, the galaxies are at a distance where there begins to
be signs of selection bias.  A detailed SBF investigation using \hst\
would finally allow the complicated structure and dynamics of this
important nearby region to be uncovered.  Finally, we note that the
zero~point of the early-type galaxy distance scale will be improved by
forthcoming \hst\ measurements of Cepheid distances to late-type galaxies
physically associated with ellipticals.  This will secure our
understanding of the physical structure of the Local Supercluster and
tighten the constraints on \ho\ from early-type galaxies.

\section*{Acknowledgments}
We thank our SBF and SMAC survey collaborators Ed Ajhar, Roger Davies,
Alan Dressler, David Schlegel, and Russell Smith for their enormous
efforts in these projects.   We are grateful for enlightening
email exchanges and encouragement from Jeff Willick early in this project.
Despite his many responsibilities and commitments, Jeff was never too
busy to share the benefits of his knowledge and experience.  This work was
supported at the University of Durham by a PPARC rolling grant in
Extragalactic Astronomy and Cosmology and made use of Starlink computer
facilities.  JPB thanks the ACS project at Johns Hopkins University
for support while finishing this paper.


\begin{table*}
\begin{minipage}{150mm}
\caption{Summary of Distance Comparison}\label{tab:distcomp}
\tabcolsep=0.35cm\small
\begin{tabular}{cccrcccc}
\hline
Comparison~  & Type & $v_{\rm max}$ & $N_g$ & 
$H_0$\footnote[1]{These values of $H_0$ are derived from the SBF-II direct
calibration, based on the Cepheid distances tabulated by Ferrarese \etal\
(2000); changing to the revised Cepheid distances of Freedman \etal\
(2001) would increase each $H_0$ by 2.8\% (see Ajhar \etal\ 2001).}
 & $\chi^2_\nu(1p)$\footnote[2]{Reduced $\chi^2$ for a single parameter ($H_0$)
fit between the two sets of distances.}
 & slope\footnote[3]{Best-fitting slope for a 2-parameter, bivariate linear fit of the
first set of distances as a function of the second set.}
 & $\chi^2_\nu(2p)$\footnote[4]{Reduced $\chi^2$ for the 2-parameter fit.} \\
\hline
FP vs SBF\dotfill   & all & \dots & 164 & $69.8\pm1.1$ & 1.23 & $1.14\pm0.05$ & 1.20 \\
FP vs SBF\dotfill   &  E  & \dots & 111 & $68.6\pm1.4$ & 0.88 & $1.05\pm0.06$ & 0.91 \\
FP vs SBF\dotfill   & all &  3000 & 133 & $68.0\pm1.3$ & 1.26 & $1.15\pm0.06$ & 1.25 \\
FP vs SBF\dotfill   &  E  &  3000 &  89 & $67.4\pm1.6$ & 0.89 & $1.00\pm0.08$ & 0.93 \\[5pt]
FP vs SBF, red\footnote[5]{`Red' subsample defined in Paper~I:
galaxies with $\viz\ge1.135\,$.}\dotfill & all & \dots & 136 & $70.3\pm1.3$ & 1.18 & $1.11\pm0.05$ & 1.17 \\
FP vs SBF, red{$^e$}\,\dotfill &  E  & \dots &  94 & $68.4\pm1.5$ & 0.80 & $1.05\pm0.06$ & 0.84 \\
FP vs SBF, red{$^e$}\,\dotfill & all &  3000 & 105 & $68.2\pm1.5$ & 1.20 & $1.10\pm0.07$ & 1.20 \\
FP vs SBF, red{$^e$}\,\dotfill &  E  &  3000 &  72 & $66.9\pm1.8$ & 0.79 & $0.98\pm0.09$ & 0.83 \\[5pt]
SBF vs \iras\dotfill & all & \dots & 164 & $75.4\pm0.9$ & 1.08 & $1.00\pm0.03$ & 1.09 \\
SBF vs \iras\dotfill &  E  & \dots & 111 & $75.3\pm1.1$ & 1.18 & $1.02\pm0.04$ & 1.20 \\
SBF vs \iras\dotfill & all &  3000 & 133 & $74.1\pm1.1$ & 1.00 & $1.13\pm0.05$ & 0.96 \\
SBF vs \iras\dotfill &  E  &  3000 &  89 & $74.2\pm1.2$ & 1.13 & $1.20\pm0.07$ & 1.04 \\[5pt]
FP vs \iras\dotfill   & all & \dots & 164 & \dots & 1.18 & $1.10\pm0.05$ & 1.16 \\
FP vs \iras\dotfill   &  E  & \dots & 111 & \dots & 1.08 & $1.07\pm0.06$ & 1.07 \\
FP vs \iras\dotfill   & all &  3000 & 133 & \dots & 1.23 & $1.32\pm0.08$ & 1.09 \\
FP vs \iras\dotfill   &  E  &  3000 &  89 & \dots & 1.13 & $1.30\pm0.11$ & 1.05 \\[5pt]
FP vs \iras, red{$^e$}\,\dots  & all & \dots & 136 & \dots & 1.11 & $1.09\pm0.05$ & 1.09 \\
FP vs \iras, red{$^e$}\,\dots  &  E  & \dots &  94 & \dots & 0.93 & $1.07\pm0.06$ & 0.93 \\
FP vs \iras, red{$^e$}\,\dots  & all &  3000 & 105 & \dots & 1.15 & $1.32\pm0.09$ & 1.02 \\
FP vs \iras, red{$^e$}\,\dots  &  E  &  3000 &  72 & \dots & 0.97 & $1.29\pm0.12$ & 0.89 \\[5pt]
\Nbar\,SBF vs \iras\,\footnote[6]{Adopts the \Nbar--distance calibration of 
Eq.\,(\ref{eq:nbarmbar}); the fact that the resulting $H_0$'s are all 1\% larger
than those from the standard SBF method implies a $\sim\,$0.02 mag offset of the
SBF-IV \Nbar\ calibration with respect to the SBF-II \viz\ calibration.}\dotfill
 & all & \dots & 164 & $76.2\pm0.9$ & 0.96 & $0.97\pm0.03$ & 0.96 \\
\Nbar\,SBF vs \iras\,{$^f$}\dotfill &  E  & \dots & 111 & $76.4\pm1.1$ & 1.07 & $0.98\pm0.04$ & 1.08 \\
\Nbar\,SBF vs \iras\,{$^f$}\dotfill & all &  3000 & 133 & $74.8\pm1.0$ & 0.87 & $1.09\pm0.05$ & 0.86 \\
\Nbar\,SBF vs \iras\,{$^f$}\dotfill &  E  &  3000 &  89 & $74.9\pm1.2$ & 1.01 & $1.15\pm0.07$ & 0.96 \\
\hline
\end{tabular}\vspace{-12pt}
\end{minipage}\end{table*}

\end{document}